\documentclass[11pt,english,onecolumn]{IEEEtran}
\usepackage[T1]{fontenc}
\usepackage[latin9]{inputenc}
\usepackage{geometry}
\usepackage{mathrsfs}
\usepackage{enumitem}
\usepackage{amsmath}
\usepackage{amsthm}
\usepackage{amssymb}
\usepackage{graphicx}
\usepackage{setspace}
\makeatletter
\theoremstyle{plain}
\newtheorem{thm}{\protect\theoremname}
\theoremstyle{definition}
\newtheorem{defn}[thm]{\protect\definitionname}
\theoremstyle{plain}
\newtheorem{lem}[thm]{\protect\lemmaname}
\theoremstyle{plain}
\newtheorem{fact}[thm]{\protect\factname}
\theoremstyle{plain}
\newtheorem{prop}[thm]{\protect\propositionname}

\usepackage{hyperref}
\usepackage{enumitem}
\usepackage{breqn}
\usepackage{bbm} 
\usepackage{cite}

 \DeclareMathOperator*{\argmin}{arg\,min}

\allowdisplaybreaks

\global\long\def\s[#1]{\textnormal{\scriptsize #1}}
\global\long\def\st[#1]{\textnormal{\tiny #1}}

\global\long\def\pe{\mathsf{pe}}

\global\long\def\P{\mathbb{P}}
\global\long\def\E{\mathbb{E}}

\global\long\def\I{\mathbbm{1}}
\global\long\def\m[#1]{\boldsymbol{#1}} % matrices, or collection of vectors

\global\long\def\r[#1]{#1}

\global\long\def\dfn{:=}

\global\long\def\trre[#1,#2]{\overset{{\scriptstyle (#2)}}{#1}} % transition explained with reason

\author{
\IEEEauthorblockN{Nir Weinberger}

\IEEEauthorblockA{The Viterbi Faculty of Electrical and Computer Engineering\\
  	    Technion - Israel Institute of Technology\\
Technion City, Haifa 3200004, Israel
} \\
\IEEEauthorblockA{nirwein@technion.ac.il}\\
}

\makeatother

\usepackage{babel}
\providecommand{\definitionname}{Definition}
\providecommand{\factname}{Fact}
\providecommand{\lemmaname}{Lemma}
\providecommand{\propositionname}{Proposition}
\providecommand{\theoremname}{Theorem}

\begin{document}

\title{Error Probability Bounds for Coded-Index DNA Storage Systems\thanks{The material in this paper was accepted in part to the International
Zurich Seminar (IZS) on Communications, Zurich, Switzerland, March
2022.}}

\maketitle
\renewcommand\[{\begin{equation}}
\renewcommand\]{\end{equation}}
\thispagestyle{empty}
\vspace{0cm}
\begin{abstract}
The DNA storage channel is considered, in which a codeword is comprised
of $M$ unordered DNA molecules. At reading time, $N$ molecules are
sampled with replacement, and then each molecule is sequenced. A coded-index
concatenated-coding scheme is considered, in which the $m$th molecule
of the codeword is restricted to a subset of all possible molecules
(an inner code), which is unique for each $m$. The decoder has low-complexity,
and is based on first decoding each molecule separately (the inner
code), and then decoding the sequence of molecules (an outer code).
Only mild assumptions are made on the sequencing channel, in the form
of the existence of an inner code and decoder with vanishing error.
The error probability of a random code as well as an expurgated code
is analyzed and shown to decay exponentially with $N$. This establishes
the importance of increasing the coverage depth $N/M$ in order to
obtain low error probability. 
\end{abstract}


\section{Introduction}

In recent years, the capability of storing information on a Deoxyribonucleic
acid (DNA) medium \cite{neiman1964some} was practically demonstrated
by a few working prototypes \cite{church2012next,goldman2013towards,grass2015robust,yazdi2015rewritable,organick2018random,bornholt2016dna}.
Based on these experimental systems, the possible impairments of this
storage channel were characterized in \cite{heckel2019characterization},
and various authors have proposed and analyzed coding methods for
this storage channel \cite{church2012next,goldman2013towards,grass2015robust,yazdi2015rewritable,kiah2016codes,erlich2017dna,sala2017exact,organick2018random,lenz2019anchor,sima2021coding,tang2021error}.
In this paper, we propose and analyze a general coding method and
a suitable decoder for this storage channel. Our analysis focuses
on error probability of such systems, and specifically on its scaling,
with respect to (w.r.t.) the parameters of the system. To facilitate
this analysis, we consider random, unstructured, codes. Nonetheless,
both the codebook ensemble and the decoder take complexity considerations
into account, as a step towards effective, practical, implementation
of such systems. 

\paragraph*{The DNA storage channel model}

Information is stored in a pool of $M$ short DNA molecules, where
each such molecule is comprised of two complementary length $L$ strands
of four nucleotides (Adenine, Cytosine, Guanine, and Thymine). The
$M$ molecules cannot be spatially ordered, and during reading are
accessed in uncontrollable manner. Specifically, the $M$ molecules
are sampled from the DNA pool $N$ times (with replacement), and each
of these sampled molecules is \emph{sequenced }in order to obtain
a vector describing the $L$ nucleotides which were synthesized to
this molecule. The set of $N$ sequenced molecules is the channel
output. The sampling mechanism leads to \emph{molecule errors}, as,
e.g. some of the $M$ molecules might not be read at all (erased).\footnote{In fact, in practice, it might be that a molecule was actually not
written at all to the pool, or was erased during storage time.} The sequencing mechanism leads to \emph{symbol} \emph{errors}, modeled
as an $L$-dimensional probability kernel $W^{(L)}$ which specifies
the probability of sequencing some $L$-symbol vector conditioned
that the information synthesized to the molecule was (possibly other)
$L$-symbol vector.\footnote{In fact, in practice, it might be that a molecule was not synthesized
to the correct sequence of nucleotides during writing, or because
it was corrupted during storage. }

\paragraph*{Capacity}

The capacity of the DNA channel was first studied in \cite{shomorony2021dna},
and later on by \cite{lenz2020achievable,lenz2019upper,weinberger2021DNA}.
These works assumed that the sequencing channel is memoryless, and
specifically, a binary symmetric channel (BSC) in \cite{shomorony2021dna,lenz2020achievable,lenz2019upper}
(with some results generalized to larger alphabet symmetric channels),
and general, possibly asymmetric, discrete memoryless channels (DMC)
in \cite{weinberger2021DNA}. A fundamental observation \cite{shomorony2021dna}
has established that the capacity is positive only when the scaling
of the molecule length is $L=\beta\log M$ with $\beta>1$. In \cite{lenz2020achievable,lenz2019upper}
it was observed that the decoder can gain from observing the same
molecule multiple times with independent sequencing ``noise'' realizations,
and so the capacity of the DNA channel is related to the \emph{binomial
(multi-draw) }channel \cite{mitzenmacher2006theory} (for BSC sequencing
channels), and more generally, to \emph{information combining} \cite{sutskover2005extremes,land2005bounds,land2006information}.
For the DNA storage channel, the impediment of achieving this capacity
is that the decoder does not know the order of the output molecules.
To resolve this, \cite{lenz2020achievable} proposed a decoder based
on (hard)\emph{ clustering}\footnote{The adjective ``hard'' is in the sense that the clustering is based
on thresholding the Hamming distance between molecules.} of the $N$ output molecules, so that with high probability the decoder
can identify the output molecules which correspond to the same input
input, and thus can exploit this multi-draw BSC to obtain increased
capacity. Nonetheless, this requirement for clustering limited the
regime of $\beta$ and the crossover probability, both for the lower
bound (achievable) \cite{lenz2020achievable} and the upper bound
(converse) \cite{lenz2019upper}. This was alleviated in \cite{weinberger2021DNA}
which considered general DMCs, and significantly improved the regime
in which the capacity bounds are valid.

\paragraph*{Motivation}

In this paper, we focus on an alternative type of schemes and analysis
for several reasons: First, the decoder in \cite{weinberger2021DNA}
is computationally intensive, in the sense the even computing the
metric for a single codeword requires maximizing over all possible
sampling events which has cardinality of $M^{N}$ (the clustering
algorithm of \cite{lenz2020achievable} has $O(N)$ metric computation
complexity, but as said, it only guaranteed to successfully operate
in a restricted regime of $\beta$ and the BSC crossover probability).
Second, the results of \cite{shomorony2021dna,lenz2020achievable,lenz2019upper,weinberger2021DNA}
all assume a memoryless sequencing channel. As surveyed in \cite{heckel2019characterization},
a practical sequencing channel might be much more involved, and include,
e.g., deletions and insertions in addition to substitution errors.
Moreover, constraints on long sequences of homopolymers or constraints
on the composition of the nucleotides types in the codeword should
practically also be taken into account \cite{immink2017design,wang2019construction}.
Third, as was established in \cite{weinberger2021DNA} for the $N=\Theta(M)$
case, the error probability is dominated by sampling events, in which
some molecules are significantly under-sampled. So, while a codeword
is comprised of a total of $ML$ symbols, the error probability decays
as $e^{-\Theta(M)}$, rather than the $e^{-\Theta(ML)}$ decay rate
expected from a code of blocklength $ML$ used over a standard DMC.
This slow decay of the error probability is significant for practical
systems of finite blocklength. Fourth, even in the memoryless setting,
if the sequencing channel is almost clean, then $C(W)\lessapprox\log|{\cal X}|$,
and in this sense, there is only a marginal gain in capacity due to
information combining according to the binomial (multi-draw) channel.
For example, if we denote the capacity of the \emph{binomial (multi-draw)
}BSC as $C_{w,d}$, where $w$ is the crossover probability and $d$
is the number of independent draws, then $C_{0.01,d}\approx(0.91,0.97,0.99)$
bits for $d=1,2,3$,\footnote{For comparison $C_{0.11,d}\approx(0.5,0.71,0.83,0.9)$ bits for $d=1,2,3$.
See also Fig. \ref{fig: Capacity difference in bounds example} in
the discussion in Sec. \ref{sec:Main-results} to follow.} and quickly saturates to its maximal value of $\log_{2}|{\cal X}|=1$
bit.

\paragraph*{Our contribution}

Accordingly, and in the spirit of \cite{lenz2020achievable}, we consider
a simple, yet general, coding method for the DNA storage channel and
analyze its error probability. The scheme follows a practical approach
to this channel \cite{grass2015robust,yazdi2015rewritable,erlich2017dna,organick2018random}
in which the lack of order of the molecules is resolved by an \emph{index}.
The simplest indexing-based scheme, is based on utilizing the first
$\log_{2}M$ bits of each DNA molecule to specify its index $m\in[M]$.
Whenever there is no noise, the decoder is able to perfectly sort
the molecules using the received index. This leads to a rate loss
of $1/\beta$, which seems to be an inherent consequence of the lack
of order of the molecules in the pool. Nonetheless, if the payload
bits (the last $(\beta-1)\log_{2}M$ bits of the molecule) of such
encoding are arbitrary, then an erroneous ordering of the molecules
can be caused by a single bit flip. This motivated explicit \emph{coded-indexing
}based schemes for noisy sequencing channels \cite{shomorony2021dna,lenz2019upper,lenz2020achieving,weinberger2021DNA,meiser2020reading}.
Here, we consider a general, \emph{non-explicit coded-indexing scheme},
in which the possible molecules of the codeword are chosen from an
inner code -- a restricted subset ${\cal B}^{(L)}\subset{\cal X}^{L}$
of all possible molecules. This inner code is further partitioned
into $M$ equal cardinality sub-codes ${\cal B}_{m}^{(L)}$, so that
the $m$th molecule of a codeword is chosen only from ${\cal B}_{m}^{(L)}$.
As before, when there are no sequencing errors, the index $m$ of
a sampled molecule is determined by the sub-code ${\cal B}_{m}^{(L)}$
it belongs to. The advantage over uncoded indexing is that the inner
code ${\cal B}^{(L)}$ also protects the index from sequencing errors.
The decoder that we consider in this paper is based on a black-box
decoder for the inner code ${\cal B}^{(L)}$. Upon sampling and sequencing
$N$ molecules, this decoder uses the inner-code decoder to \emph{independently}
decode each of the sequenced molecule to a valid sequence in ${\cal B}^{(L)}$.
Since the decoder operates on a molecule-by-molecule basis, it is
fairly practical ($L=\beta\log M$ is expected to be relatively short,
say on the order of $10^{2}-10^{3}$), and much simpler than the decoder
of \cite{weinberger2021DNA}.\footnote{The clustering decoder \cite{lenz2020achieving} is based on sequential
assignment of molecules to clusters and thus the complexity of computing
the metric is $\Theta(N)$, but there are no guarantees on the decay
rate of the error probability, as we show in this paper.} 

After the individual molecule decoding stage, the decoder holds $N$
sequences from ${\cal B}^{(L)}$, which are partitioned to the $m$
sub-codes ${\cal B}_{m}^{(L)}$. For each $m\in[M]$, the decoder
collects the set of inner-code decoded output molecules which belong
to ${\cal B}_{m}^{(L)}$ (if there are any), and either chooses from
this sub-code a unique molecule, or declares an erasure of the $m$th
molecule. An outer-code -- which restricts the possible sequences
of molecules -- is then used to correct possible erasures or undetected
erroneous molecules. Specifically, this can be achieved by a simple
minimum Hamming distance (on a molecule level) decoder.\footnote{Practically, by a \emph{maximum distance separable }code \cite{lenz2020achievable}.}
Therefore, the proposed coded-index based scheme is practical, and
its analysis is general, in the sense that very little is assumed
on the sequencing channel. It is only required that a proper decoder
for an inner code of blocklength $L$ exists, such that the error
probability decays to zero with increasing $L$. This addresses the
first two issues raised above. 

As was shown in \cite{weinberger2021DNA}, and will also be apparent
from the analysis in this paper, it appears that sequencing errors
affect the error probability to much less extent compared to unfavorable
sampling events. Indeed, bad sampling events are the origin of the
slow $e^{-\Theta(M)}$ decay of the error probability obtained in
\cite{weinberger2021DNA}, which assumed $N=\alpha M$ for some fixed
$\alpha>1$. The only way to ameliorate this behavior is by increasing
$N$. In accordance, we consider in this paper the scaling $N=\alpha_{M}M$
where $\alpha_{M}$ may either be a constant (as in \cite{shomorony2021dna,lenz2019upper,lenz2020achievable,weinberger2021DNA})
or an increasing function of $M$ (though at a rather slow rate).
As we shall see, this has a profound effect on the scaling of the
decay of the error probability. This addresses the third issue raised
above. Furthermore, if indeed the increased capacity of the binomial
(multi-draw) capacity is marginal, then increasing $N$ is not useless,
since it has an important role in determining the error probability.
This addresses the fourth issue discussed above. Our main result pertains
to an achievable single-letter bound on the error probability of a
coded-index based scheme for the DNA storage channel. It is comprised
of both a random coding bound, as well as an expurgated bound, which
lead to a decay rate of the error probability of the order $e^{-\Theta(N)}$.
An important consequence of this result is that operating at a large
covering depth $N/M$ is of importance when low error probability
is of interest. This is in contrast to capacity analysis, which, as
discussed in \cite[Sec. I]{shomorony2021dna}, it is wasteful to operate
at a large covering depth $N/M$ as this only provides marginal capacity
gains, but the sequencing costs are large. 

\paragraph*{Comparison with previous works}

In \cite{lenz2020achievable} a coding scheme was proposed which is
similarly based on (explicit) coded-indexing and on concatenated coding
of outer code aimed to correct erasures, and an inner code aimed to
correct sequencing errors. The main difference is, perhaps, that \cite{lenz2020achievable}
aims to achieve the capacity of the multi-draw channel. To this end,
it should be recalled that the encoder does not know in advance how
many times each molecule will be sampled, and thus also not the capacity
of the effective channel from the input molecule to the corresponding
output molecule(s). The scheme \cite{lenz2020achievable} incorporates
a code over multiple molecules, so that the capacity of the jointly
encoded molecules is averaged over the randomness of the sampling
operation. Compared to this paper, \cite{lenz2020achievable} is based
on (hard) output clustering, and so is mainly tailored to the BSC,
and positive rate is only achieved when the crossover probability
is less than $1/4$. The resulting capacity lower bound tends to capacity
bound of \cite{lenz2019upper} only if the crossover probability of
the BSC tends to zero (as in our scheme, as the increase in capacity
due to multi-draws is marginally small), or if $N/M\to\infty$ (which
is indeed better than our scheme in terms of rate, but the improvement
in the error probability is not established). As described above,
in this paper, we consider general sequencing channels, and focus
on error probability analysis and simple decoding, at the price of
possible rate loss. We also mention that \cite{kovavcevic2018codes}
have studied the coding rate loss of explicit uncoded indexing compared
to optimal codes, under an adversarial channel model (with a different
scaling of the molecule size). In our context, the conclusion is that
the loss is more profound for small $\beta$. 

\paragraph*{Outline}

The rest of the paper is organized as follows. In Sec. \ref{sec:Problem-Formulation}
we establish notation conventions, and formulate the DNA storage channel
and coded-index based systems. In Sec. \ref{sec:Main-results} we
state our main result, and in Sec. \ref{sec:Proof-of-Theorem} we
provide the proof, which includes both random coding analysis as well
as an expurgated bound. In Sec. \ref{sec:Summary} we summarize the
paper. 

\section{Problem Formulation \label{sec:Problem-Formulation}}

\subsection{Notation Conventions}

Random variables will be denoted by capital letters, specific values
they may take will be denoted by the corresponding lower case letters,
and their alphabets will be denoted by calligraphic letters. Random
vectors and their realizations will be super-scripted by their dimension.
For example, the random vector $A^{K}=(A_{0},\ldots,A_{K-1})\in{\cal A}^{K}$
(where $K\in\mathbb{N}^{+}$), may take a specific vector value $a^{K}=(a_{0},\ldots,a_{K-1})\in{\cal A}^{K}$,
the $K$th order Cartesian power of ${\cal A}$, which is the alphabet
of each component of this vector. The Cartesian product of ${\cal A}$
and ${\cal B}$ (both finite alphabets) will be denoted by ${\cal A}\times{\cal B}$.
The probability of the event ${\cal {\cal E}}$ will be denoted by
$\P({\cal {\cal E}})$, and its indicator function will be denoted
by $\I({\cal E})$. The expectation operator w.r.t. a given distribution
$P$ will be denoted by $\E[\cdot]$. 

Logarithms and exponents will be understood to be taken to the natural
base. The binary entropy function $h_{b}\colon[0,1]\to[0,1]$ will
be denoted by $h_{b}(a)\dfn-a\log a-(1-a)\log(1-a)$ and the binary
Kullback--Leibler (KL) divergence $d_{b}\colon[0,1]\times(0,1)\to\mathbb{R}^{+}$
by $d_{b}(a||b)\dfn a\log\frac{a}{b}+(1-a)\log\frac{(1-a)}{(1-b)}.$ 

The number of \emph{distinct} elements of a finite multiset ${\cal A}$
will be denoted by $|{\cal A}|$. The equivalence relation will be
denoted by $\equiv$, and will mainly be used to simplify notation
at some parts of the paper (typically, the removal of subscripts/superscripts
in order to avoid cumbersome notation). Asymptotic Bachmann--Landau
notation will be used. Specifically, for a pair of positive sequences
$\{f_{K}\}_{K\in\mathbb{N}},\{g_{K}\}_{K\in\mathbb{N}}$ $f_{K}=O(g_{K})\Leftrightarrow\limsup_{K\to\infty}\frac{f_{K}}{g_{K}}<\infty$,
and $f_{K}=\Theta(g_{K})\Leftrightarrow\{f_{K}=O(g_{K})\text{ and }g_{K}=O(f_{K})\}$,
$f_{K}=o(g_{K})\Leftrightarrow\lim_{K\to\infty}\frac{|f_{K}|}{g_{K}}=0$,
and $f_{K}=\omega(g_{K})\Leftrightarrow\lim_{K\to\infty}\frac{|f_{K}|}{g_{K}}=\infty$.
Minimum and maximum will be denoted as $\min(a,b)\dfn a\wedge b$,
$\max(a,b)\dfn a\vee b$, and $[a]_{+}\dfn a\vee0$. For a positive
integer $N$, $[N]\dfn\{0,1,\ldots,N-1\}$, where scalar multiplications
of these sets will be used, e.g., as $\frac{1}{N}[N+1]=\{0,\frac{1}{N},\ldots\frac{N-1}{N},1\}$.
Throughout, for the sake of brevity, integer constraints on large
numbers which are inconsequential will be ignored, for example, the
number of codewords in a rate $R$ codebook of dimension $K$ will
be simply written as $e^{KR}$ (instead of $\lceil e^{KR}\rceil$). 

\subsection{Formulation of the DNA Storage Channel \label{subsec:Formulation-of-the}}

In this section, we formulate a DNA storage channel and the corresponding
encoders and decoders. The channel will be indexed by the number of
molecules $M$ in the DNA pool used to store a message.\footnote{Occasionally, also by quantities which depend on $M$. Specifically,
we may index the channels by $N\equiv N_{M}$, the number of output
molecules.}

\paragraph*{The channel model (reading mechanism)}

A DNA molecule is a sequence of $L\equiv L_{M}\in\mathbb{N}_{+}$
nucleotides (symbols) chosen from an alphabet ${\cal X}$, where in
physical systems ${\cal X}=\{\mathsf{A},\mathsf{C},\mathsf{G},\mathsf{T}\}$.
In previous works (e.g., \cite{shomorony2021dna,lenz2019upper,lenz2020achievable})
the binary case ${\cal X}=\{0,1\}$ was typically assumed for simplicity,
yet here, we do not make any such assumptions. Thus, each molecule
is uniquely represented by a sequence $x^{L}\in{\cal X}^{L}$. An
input to the DNA channel is a sequence of $M$ molecules, $x^{LM}=(x_{0}^{L},\ldots x_{M-1}^{L})$,
where $x_{m}^{L}\in{\cal X}^{L}$ for all $m\in[M]$.\footnote{In principle, a codeword is a \emph{multiset }of $M$ molecules, that
is, the order is not specified. However, for analysis it is convenient
to assume an arbitrary ordering of the molecules. As evident from
the description of the sampling stage, this ordering does not affect
the channel output.} 

Suppose that an input is synthesized into a sequence of $M$ molecules,
$x^{LM}$. The DNA storage channel we consider here is parameterized
by the number of molecule samples $N\equiv N_{M}\in\mathbb{N}_{+}$,
and a sequencing channel $W^{(L)}\colon{\cal X}^{L}\to{\cal Y}^{(L)}$.
Note that the output alphabet ${\cal Y}^{(L)}$ does not have to be
the $L$th order Cartesian power of ${\cal Y}$. For example, for
a deletion/insertion sequencing channel ${\cal Y}^{(L)}={\cal Y}^{0}\cup{\cal Y}^{1}\cup\cdots$. 

The operation of the channel on the stored codeword $x^{LM}$ is modeled
as a two-stage process:
\begin{enumerate}
\item Sampling: $N$ molecules are sampled uniformly from the $M$ molecules
of $x^{LM}$, independently, with replacement. Let $U^{N}\in[M]^{N}$
be such that $U_{n}$ is the sampled molecule at sampling event $n\in[N]$.
We refer to $U^{N}$ as the \emph{molecule index vector}, so that
$U^{N}\sim\text{Uniform}([M]^{N})$. The result of the sampling stage
is thus the vector
\[
(x_{U_{0}}^{L},x_{U_{1}}^{L},\ldots,x_{U_{N-1}}^{L})\in({\cal X}^{L})^{N}.
\]
Let $S^{M}\in[N]^{M}$ be such that $S_{m}$ is the number of times
that molecule $m$ was sampled, to wit $S_{m}=\sum_{n\in[N]}\I\{U_{n}=m\}$,
the empirical count of $U^{N}$. It holds that $S^{M}\sim\text{Multinomial}(N;(\frac{1}{M},\frac{1}{M},\ldots\frac{1}{M}))$,
and we refer to $S^{M}$ as the \emph{molecule duplicate vector}. 
\item Sequencing: For each $n\in[N]$, $x_{U_{n}}^{L}$ is sequenced to
$Y_{n}^{(L)}\in{\cal Y}^{(L)}$, and the sequencing of $x_{U_{n}}^{L}$
is independent for all $n\in[N]$. Denoting the channel output by
$(Y^{(L)})^{N}=(Y_{0}^{(L)},\ldots,Y_{N-1}^{(L)})\in({\cal Y}^{(L)})^{N}$,
it thus holds that 
\[
\P\left[(Y^{(L)})^{N}=(y^{(L)})^{N}\mid x^{LM},\;U^{N}\right]=\prod_{n\in[N]}W^{(L)}\left(y_{n}^{(L)}\mid x_{U_{n}}^{L}\right).
\]
\end{enumerate}
The channel output is $(Y^{(L)})^{N}$, where due to the random sampling
stage, it is clear that the observed index $n$ of $y_{n}^{(L)}$
in $(y^{(L)})^{N}$ is immaterial for decoding.

\paragraph*{The encoder}

A codebook to the DNA storage channel is a set of different possible
codewords (channel inputs) ${\cal C}=\{x^{LM}(j)\}$. In this work
we consider the following restricted set of codebooks, which is based
on \emph{coded}-\emph{index}:
\begin{defn}[Coded-index based codebooks]
\label{def: index based coding }Let $\{{\cal B}_{m}^{(L)}\}_{m\in[M]}$
be a collection of pairwise disjoint sets ${\cal B}_{m}^{(L)}\subset{\cal X}^{L}$
of equal cardinality, and let ${\cal B}^{(L)}=\cup_{m\in[M]}{\cal B}_{m}^{(L)}$
be their union. A DNA storage code is said to be \emph{coded-index
}based if $x_{m}^{L}(j)\in{\cal B}_{m}^{(L)}$ for all $m\in[M]$
and all $j\in[|{\cal C}|]$. 
\end{defn}
To wit, a codeword contains exactly a single molecule from each of
the $M$ sub-codes $\{{\cal B}_{m}^{(L)}\}_{m\in[M]}$. The identity
of the sub-code from which $x_{m}^{L}(j)$ was chosen from is considered
an ``index'' of the molecule, which can be used by the decoder to
order the decoded molecules. A coded-index based codebook can be thought
of as a concatenated code. The set ${\cal B}^{(L)}$ is an inner-code,
which is used to clean the output molecules from sequencing errors,
and the dependency between molecules of different index $m$ can be
considered an outer-code which is used to cope with erasures due to
the sampling stage, and residual sequencing errors. 

\paragraph*{The decoder}

A general decoder is a mapping $\mathsf{D}\colon({\cal Y}^{(L)})^{N}\to[|{\cal C}|]$.
In this work we consider the following class of decoders, which are
suitable for coded-index based codebooks. The decoder is equipped
with an inner-code decoder $\mathsf{D}_{b}\colon{\cal Y}^{(L)}\to{\cal B}^{(L)}$
and a threshold $T\equiv T_{M}$, and processes $(y^{(L)})^{N}$ in
three steps:
\begin{enumerate}
\item Correction of individual molecules: The decoder employs the inner-code
decoder for each of the received molecules $y_{n}^{(L)}$, $n\in[N]$,
and sets $z_{n}^{L}=\mathsf{D}_{b}(y_{n}^{(L)})$. Following this
stage, it holds that $z^{LN}=(z_{0}^{L},\ldots,z_{N-1}^{L})$ is such
that $z_{n}^{L}\in{\cal B}^{(L)}$ for all $n\in[N]$. 
\item Threshold for each index: For each index $m\in[M]$, if there exists
a $b^{L}\in{\cal B}_{m}^{(L)}$ such that 
\begin{equation}
\sum_{n\in[N]}\I\{z_{n}^{L}=b^{L}\}\geq T>\max_{\tilde{b}^{L}\in{\cal B}_{m}^{(L)}\backslash\{b_{l}^{L}\}}\sum_{n\in[N]}\I\{z_{n}^{L}=\tilde{b}^{L}\}\label{eq: decoder threshold definition}
\end{equation}
then the decoder sets $\hat{x}_{m}^{L}=b^{L}$. That is, $\hat{x}_{m}^{L}=b^{L}$
if $b^{L}$ is a unique molecule in ${\cal B}_{m}^{(L)}$ whose number
of appearances in $z^{LN}$ is larger than $T$. Otherwise $\hat{x}_{m}^{L}=\mathsf{e}$,
where $\mathsf{e}$ is a symbol representing an \emph{erasure}. 
\item Codeword decoding: Let 
\[
j^{*}=\argmin_{j\in[|{\cal C}|]}\rho\left(\hat{x}^{LM},x^{LM}(j)\right),
\]
where (with a slight abuse of notation)
\begin{equation}
\rho(\hat{x}^{LM},x^{LM})\dfn\sum_{m\in[M]}\rho(\hat{x}_{m}^{L},x_{m}^{L}),\label{eq: definition of rho distance}
\end{equation}
and 
\[
\rho(\hat{x}^{L},x^{L})\dfn\begin{cases}
\I\{\hat{x}^{L}\neq x^{L}\}, & \hat{x}^{L}\neq\mathsf{e}\\
0, & \hat{x}^{L}=\mathsf{e}
\end{cases}.
\]
That is, the distance of a codeword $x^{LM}(j)$ to $\hat{x}^{LM}$
has zero contribution from erased molecule indices or if $x_{m}^{L}(j)=\hat{x}_{m}^{L}$,
and $1$ otherwise. 
\end{enumerate}

\paragraph*{Assumptions on the channel model}
\begin{enumerate}
\item Molecule length scaling: $L\equiv L_{M}=\beta\log M$, where $\beta>1$
is the \emph{molecule length parameter}.
\item Coverage depth scaling: $N=\alpha_{M}M$, where $\alpha_{M}>1$ is
the \emph{coverage depth scaling function}, and $\alpha_{M}$ is a
monotonic non-decreasing function. If $\alpha\equiv\alpha_{M}$ is
constant then $\alpha$ is the \emph{coverage depth scaling parameter.} 
\end{enumerate}
The DNA storage channel is thus indexed by $M$ and parameterized
by $(\alpha_{M},\beta,\{W^{(L)}\}_{L\in\mathbb{N}_{+}})$. The (storage)
rate of the codebook ${\cal C}$ is given by $R=\frac{\log|{\cal C}|}{ML}$,
and the error probability of $\mathsf{D}$ given that $x^{LM}(j)\in{\cal C}$
was stored is given by 
\begin{equation}
\pe({\cal C},\mathsf{D}\mid x^{LM}(j))\dfn\P\left[\mathsf{D}((y^{(L)})^{N})\neq j\mid x^{LM}(j)\right].\label{eq: error proabability}
\end{equation}
Let $\psi_{M}\colon\mathbb{N}_{+}\to\mathbb{N}_{+}$ be a monotonic
strictly increasing sequence. An \emph{error exponent $E(R)$ w.r.t.
scaling $\psi{}_{M}$ }is achievable for channel $\mathsf{DNA}$ at
rate $R$, if there exists a sequence $\{{\cal C}_{M},\mathsf{D}_{M}\}_{M\in\mathbb{N}_{+}}$so
that the average error probability is bounded as
\[
\liminf_{M\to\infty}-\frac{1}{\psi_{M}}\log\left[\frac{1}{|{\cal C}_{M}|}\sum_{j\in[|{\cal C}_{M}|]}\pe({\cal C}_{M},\mathsf{D}_{M}\mid x^{LM}(j))\right]\geq E(R).
\]
In this paper, we present single-letter expressions for error exponents
achieved under coded-index codebook and the class of decoders defined
above. Let $R_{b}=\frac{\log|{\cal B}^{(L)}|}{L}$ be the rate of
the inner code, and let 
\begin{equation}
\pe_{b}({\cal B}^{(L)})=\max_{b^{L}\in{\cal B}^{(L)}}W^{(L)}\left[\mathsf{D}_{b}(y^{(L)})\neq b^{L}\mid b^{L}\right]\label{eq: inner code maximal error probability}
\end{equation}
be the maximal error probability of the inner code over the sequencing
channel $W^{(L)}$. We will not make any assumptions regarding the
channel $W^{(L)}$ (e.g., that it is the $L$th order power of a DMC
$W\colon{\cal X}\to{\cal Y}$, or that $W^{(L)}$ is a deletion/insertion
channel). Rather instead, we will assume that a suitable sequence
of codes can be found, as follows. 

\paragraph*{Assumptions on the inner code}
\begin{enumerate}
\item Inner code rate: $R_{b}=\frac{1}{L}\log|{\cal B}^{(L)}|>1/\beta$.
\item Vanishing inner code error probability: $\pe_{b}({\cal B}^{(L)})=e^{-\Theta(L^{\zeta})}$
where $\zeta>0$. 
\end{enumerate}
The cardinality of each sub-code is $|{\cal B}_{m}^{(L)}|=\frac{e^{R_{b}L}}{M}=M^{(R_{b}\beta-1)}$.
Thus, for any rate which satisfies the assumption $R_{b}>1/\beta$,
the inner sub-code size $|{\cal B}_{m}^{(L)}|$ is strictly larger
than $1$ for all $M$ large enough. The assumption on the error probability
assures that the error probability at the first decoding step tends
to zero as $L=\beta\log M\to\infty$. Thus, it must hold that $R_{b}$
is below the normalized capacity of the sequencing channel $\lim_{L\to\infty}\frac{1}{L}C(W^{(L)})$,
where, in general, the capacity is as given by the \emph{infimum information
rate} formula of \cite{verdu1994general}. For memoryless sequencing
channels $W^{(L)}=W^{\otimes L}$, and the error probability decays
exponentially with $L$, as $e^{-E\cdot L}$ where $E$ is an exponent
which depends on the rate of the inner code. Here, however, we consider
general sequencing channels, for which the error probability may decay
much more slowly with $L$, even for optimal codes. For concreteness,
we have assumed $\pe_{b}({\cal B}^{(L)})=e^{-\Theta(L^{\zeta})}$,
but as we shall see, $\zeta$ does not affect the achievable exponent,
and, in fact, the assumption $\pe_{b}({\cal B}^{(L)})=o(1)$ suffices
for our next result to hold (but makes the derivations in the proof
slightly more cumbersome). Moreover, we do not specify the inner code
and the decoder, and they may correspond to any suitable choice. For
example, \emph{polar codes} can be used, which can achieve error scaling
of $e^{-\Theta(\sqrt{N})}$ for standard DMCs \cite{arikan2009rate,hassani2012rate},
and of $e^{-\Theta(N^{1/3})}$ for channels which include insertions,
deletions, and substitutions \cite{tal2019polar}. 

In order to derive achievable error probability bounds, we will consider
the following random coding ensemble.
\begin{defn}[Coded-index based random coding ensemble]
\label{def:RCE}For a given $\{{\cal B}_{m}^{(L)}\}_{m\in[M]}$ (with
the notation of Definition \ref{def: index based coding }), let ${\cal C}=\{X^{LM}(j)\}$
be a random code such that $X_{m}^{L}(j)$ is chosen uniformly at
random from ${\cal B}_{m}^{(L)}$, independently for all $m\in[M]$
and all $j\in[|{\cal C}|]$.
\end{defn}

\section{Main Result \label{sec:Main-results}}

Our main result is as follows:
\begin{thm}
\label{thm: main achievable bound}Let $m$ sub-codes be given $\{{\cal B}_{m}^{(L)}\}_{m\in[M]}$,
and let $\mathsf{D}_{b}$ be a decoder which satisfy the assumptions
on the inner code for ${\cal B}^{(L)}=\cup_{m\in[M]}{\cal B}_{m}^{(L)}$
($R_{b}>1/\beta$, $\pe_{b}({\cal B}^{(L)})=e^{-\Theta(L^{\zeta})}$).
Then, there exists a sequence of codebooks ${\cal C}_{M}$ and corresponding
threshold-based decoders (as described in Sec. \ref{subsec:Formulation-of-the})
so that: 
\begin{itemize}
\item If $N/M=\Theta(1)$ then 
\[
\liminf_{M\to\infty}-\frac{1}{M}\log\pe({\cal C}_{M},\mathsf{D}_{M})\geq d_{b}\left(1-\frac{R}{R_{b}-1/\beta}\,\middle\vert\middle\vert\,e^{-\frac{N}{M}}\right)
\]
for any $R<(R_{b}-1/\beta)(1-e^{-\frac{N}{M}})$. 
\item If $N/M=\omega(1)$ then 
\[
\liminf_{N\to\infty}-\frac{1}{N}\log\pe({\cal C}_{M},\mathsf{D}_{M})\geq\begin{cases}
\frac{1}{2}\left[1-\frac{R}{R_{b}-1/\beta}\right], & \frac{N}{ML}<2(R_{b}-1/\beta)\\
\frac{ML}{N}\left[R_{b}-1/\beta-R\right], & 2(R_{b}-1/\beta)\leq\frac{N}{ML}<4(R_{b}-1/\beta)\\
\frac{1}{4}\left[1-\frac{R}{R_{b}-1/\beta}\right], & \frac{N}{ML}>4(R_{b}-1/\beta)
\end{cases}
\]
for any $R<R_{b}-1/\beta$.
\end{itemize}
\end{thm}
The proof of Theorem \ref{thm: main achievable bound} follows directly
by taking the better of a random coding bound (Prop. \ref{prop: index molecule random coding bound}
in Sec. \ref{subsec:Random-Coding-Analysis}) and an expurgated bound
(Prop. \ref{prop: index molecule expurgated bound} in Sec. \ref{subsec:Expurgated-Bound-Analysis}).

\paragraph*{Discussion}
\begin{enumerate}
\item The exponent bound is not continuous in $N$ (that is, there is a
phase transition), and the behavior is markedly different between
$N=\Theta(M)$ and $N=\omega(M)$. As emanates from the analysis,
in both regimes, the threshold is chosen so that $T\equiv T_{M}=o(N/M)$.
This is because the error probability of the inner code decay to zero
as $e^{-\Theta(L^{\zeta})}=e^{-\Theta(\log^{\zeta}M)}$, and so the
average number of erroneously sequenced molecules is $o(N/M)$ per
molecule. It is plausible that similar bounds can be obtained with
a simpler decoder which only checks if there exists a unique molecule
of index $m$, and otherwise declares an erasure (which is equivalent
to setting $T=1$), or a decoder based on majority for each index
$m$. Nonetheless, the insight from the random coding analysis of
our general threshold decoder is that even if $N/M$ is large, there
is no gain in setting the threshold to be $\Theta(N/M)$. This is
not obvious \emph{a priori}. 
\item The only regime in which the expurgated bound is better than the random
coding bound is $\frac{N}{ML}>4(R_{b}-1/\beta)$. Consequently, in
the two other regimes the error probability bound holds for a typical
code from the ensemble. 
\item The result does not depend on $\zeta$, the assumed scaling of the
inner code error probability ($\pe_{b}({\cal B}^{(L)})=e^{-\Theta(L^{\zeta})}$),
and manifests the fact that sampling events dominate the error probability,
compared to sequencing error events. 
\item For the standard channel coding problem over DMCs with blocklength
$N$, the method of types leads to random coding and expurgated bounds
which tend to their asymptotic values up to a $O((\log N)/N)$ term
(this can be avoided using Gallager's method \cite[Ch. 5]{gallager1968information},
see also \cite[Problem 10.33]{csiszar2011information}). Here, it
is evident from the proof that the decay is much slower, and could
be as slow as $O(1/\log M)$. As discussed in \cite[Sec. VII]{weinberger2021DNA}
this seems to be an inherent property of this channel.
\item Proof outline: The proof appears in Sec. \ref{sec:Proof-of-Theorem}
that follows. There, in Sec. \ref{subsec:Error-Events-for}, we analyze
the channel operation, and evaluate the probability of events in which
either some of the molecules are not sampled enough times, or that
there are many sequencing errors (respectively, Lemmas \ref{lem:Bounds on the size of erasure and undetected sets},
\ref{lem: amplification large deviations} and \ref{lem: total seqeuncing errors large deviations}).
Using these bounds, we evaluate a large-deviation bound for the number
of erasures and undetected errors in the output of the decoder (Lemma
\ref{lem: erausre large deviations with high threshold}). Then, in
Sec. \ref{subsec:Random-Coding-Analysis} we consider the random coding
ensemble of Definition \ref{def:RCE}, and evaluate the probability
that a single randomly chosen codeword from this ensemble is decoded
instead of the true codeword, conditioning on a specific number of
erasures and undetected errors (Lemma \ref{lem: random coding pairwise error probability}).
This bound uses the erasure/undetected-error probability bounds of
Sec. \ref{subsec:Error-Events-for}. A clipped union bound is then
invoked to prove the random coding bound. In Sec. \ref{subsec:Expurgated-Bound-Analysis},
the expurgated bound is proved by first establishing the existence
of a codebook with specific values of distance spectrum, according
to the $\rho$ distance defined in \eqref{eq: definition of rho distance}
(Lemma \ref{lem: distance spectrum expurgation}). The expurgated
bound is then proved by upper bounding the error probability of this
codebook by the sum of all its pairwise error probabilities. 
\item Theorem \ref{thm: main achievable bound} only provides an achievable
bound, and establishing tightness of this bound seems to be a challenging
task for the multinomial model assumed here. Recall that the sampling
model is such that each molecule from the codeword $x^{LM}$ is chosen
uniformly at random, with replacement. Consequently, the molecule
duplicate vector $S^{M}$ follows a multinomial distribution. Lemma
\ref{lem: amplification large deviations} bounds the probability
that $\sigma M$ molecules have been under-sampled during the sampling
stage. The direct analysis of this event under the multinomial distribution
of $S^{M}$ is difficult. To circumvent this, the proof of the lemma
utilizes the \emph{Poissonization of the multinomial} effect, proposed
in \cite{lenz2020achieving,shomorony2021dna} to analyze the capacity
of the channel. Specifically, the aforementioned probability is upper
bounded by the probability of the same sampling event, while replacing
$S^{M}$ with an independent and identically distributed (i.i.d.)
Poisson r.v.'s, $\tilde{S}_{m}\sim\text{Pois}(N/M)$, $m\in[M]$.
While this provably upper bounds the error probability, this upper
bound is tight at the center of the multinomial distribution, but
may be loose at its tails. It is thus anticipated that a lower bound
on the error probability would require a direct analysis of the multinomial
distribution. Nonetheless, if one adopts a pure Poisson model for
the number of times each molecule is sampled, i.e., $S_{m}\sim\text{Pois}(N/M)$
i.i.d., then a simple lower bound can be derived using the following
considerations. The probability that a molecule is erased is at least
$\P[\text{Pois}(N/M)\leq T]$, and is independent over different molecules.
Thus, the error probability of the code is larger than that of the
optimal error probability in an erasure channel with erasure probability
$\P[\text{Pois}(N/M)\leq T]$, neglecting undetected errors, which
will only increase the error probability (informally speaking). The
optimal error exponent for binary erasure channel can then be lower
bounded by standard bounds for erasure channel, e.g., the sphere packing
bound \cite[Thm. 5.8.1]{gallager1968information}, \cite[Thm. 10.3]{csiszar2011information}.
\item For $N/M=\Theta(1)=\alpha$, and a DMC sequencing channel $W$, a
lower bound on capacity (which is tight for all $\beta$ above a critical
value depending on $W$), was obtained in \cite[Thm. 5]{weinberger2021Error}.
This lower bound is given by 
\[
\max_{P_{X}\in{\cal P}({\cal X})}\sum_{d\in\mathbb{N}^{+}}\pi_{\alpha}(d)\cdot I(P_{X},W^{\oplus d})-\frac{1}{\beta}\left(1-\pi_{\alpha}(0)\right),
\]
where $\pi_{\alpha}(d)\dfn\frac{\alpha^{d}e^{-\alpha}}{d!}$ for $d\in\mathbb{N}$
is the Poisson probability mass function for parameter $\alpha$,
$I(P_{X},V)$ is the mutual information of a channel $V\colon{\cal A}\to{\cal B}$
with input distribution is $P_{X}$, and $V^{\otimes d}$ is the \emph{$d$-order
binomial extension of $V$}, that is the DMC $V^{\oplus d}\colon{\cal A}\to{\cal B}^{d}$
for which 
\[
V^{\oplus d}[b^{d}\mid a]=\prod_{i=0}^{d-1}V(b_{i}\mid a).
\]
From Theorem \ref{thm: main achievable bound}, a lower bound on the
maximal rate achieved by the considered scheme is $(R_{b}-1/\beta)(1-\pi_{\alpha}(0))$.
To maximize this lower bound, we choose the maximal possible rate
for which vanishing inner code error probability can be attained,
that is $R_{b}=I(P_{X},W)$. Thus, the difference between the capacity
lower bounds is 
\begin{equation}
\sum_{d\in\mathbb{N}^{+}}\pi_{\alpha}(d)\cdot I(P_{X},W^{\oplus d})-I(P_{X},W)\left(1-\pi_{\alpha}(0)\right).\label{eq: difference between optimal capacity and threshold capacity}
\end{equation}
This is expected since our decoder does not combines information from
multiple output molecules. The difference \eqref{eq: difference between optimal capacity and threshold capacity}
for a BSC sequencing channel with crossover probability $w$, and
an optimal input distribution $P_{X}=(1/2,1/2)$ is shown in Fig.
\ref{fig: Capacity difference in bounds example}, as a function of
$\alpha$, for various values of $w$. As expected, for a low values
of $w=10^{-3}$, this difference is rather small (at most $\sim0.01$
nats). While the difference is increasing with $w$, it saturates
as a function of $\alpha$, and can still be low for low values of
$\alpha$ (close to $1$). Note also that the difference in \eqref{eq: difference between optimal capacity and threshold capacity}
is independent of $\beta$. For the case $N/M=\omega(1)$ the lower
bound on capacity implied by Theorem \ref{thm: main achievable bound}
is $R_{b}-1/\beta$, where $R_{b}$ can be chosen to be as large as
the normalized capacity of the sequencing channel. Thus, the result
agrees with the capacity lower bound of the $N/M=\Theta(1)$ case,
when taking the limit $N/M\to\infty$. 
\begin{figure}
\centering{}\includegraphics[scale=0.3]{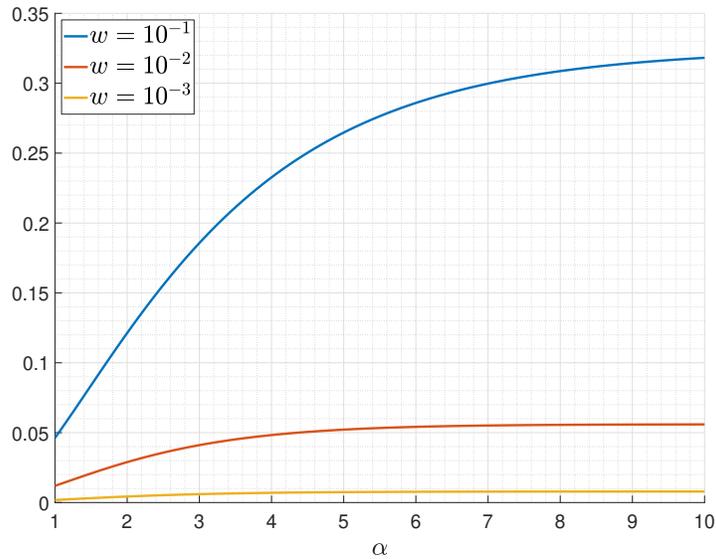}\caption{The difference between the lower bound on capacity for optimal decoding
of \cite[Thm. 5]{weinberger2021DNA} and the lower bound implied by
Theorem \ref{thm: main achievable bound} for $N/M=\alpha$, assuming
a BSC sequencing channel with crossover probability $w$ (in nats).
\label{fig: Capacity difference in bounds example}}
\end{figure}
\end{enumerate}

\section{Proof of Theorem \ref{thm: main achievable bound} \label{sec:Proof-of-Theorem}}

\subsection{Error Events for the Threshold Based Decoder \label{subsec:Error-Events-for}}

In the coded-index based coding, each codeword $x^{LM}(j)$ contains
exactly a single molecule from each of the sub-codes ${\cal B}_{m}^{(L)}$.
The molecule $x_{m}^{L}(j)$ is sampled $s_{m}$ times, where $s^{M}\in[N]^{M}$
is the molecule duplicate vector. According to the definition of the
decoder, an error in the ``inner'' decoding of the $m$th molecule,
that is $\hat{x}_{m}^{L}\neq x_{m}(j)$, may occur for several reasons.
First, it may occur that the number of times that $x_{m}(j)$ was
sampled is below the required threshold $T$, that is, the event $\I\{s_{m}\leq T\}$.
This is an erasure/undetected-error event caused by non-ideal sampling.
Second, it may occur that $x_{m}(j)$ was sampled more than $T$ times,
however, sequencing errors have caused the number of appearances of
$x_{m}(j)$ in $z^{LN}$ to drop below the threshold $T$. Third,
it may occur that $x_{m}(j)$ appears in $z^{LN}$ more than $T$
times, yet sequencing errors have caused a different molecule $\tilde{x}^{L}\in{\cal B}_{m}^{(L)}$
to appear in $z^{LN}$ more than $T$ times. If neither of these events
has occurred then the molecule $\hat{x}_{m}^{L}=x_{m}(j)$ is decoded
correctly. Otherwise, the $m$th molecule is either erased in the
second step of the decoder or is erroneously decoded (which is an
\emph{undetected error}). 

For such events to have an effect on the exponential decrease of the
error probability after the decoding of the outer code, they must
have $\Theta(M)$ occurrences. In accordance, we analyze in this section
the large deviations of the erasure-probability and undetected-error-probability.
To this end, we recall that the decoder $\mathsf{D}$ is defined by
a threshold $T$, and we next parameterize this threshold by a parameter
$\tau\in(0,\frac{1}{2})$, and write 
\begin{equation}
T\equiv T_{\tau}\dfn\frac{N}{M}\left(1-\sqrt{2\tau}\right).\label{eq: threshold definition}
\end{equation}
Note that the constraint $\tau<\frac{1}{2}$ assures that $T_{\tau}>0$.
For the analysis, we will use the following notation for random variables
that correspond to channel events:
\begin{itemize}
\item $K_{m}\in[s_{m}+1]$ is the number of copies of the molecule $x_{m}^{L}(j)$
that have been erroneously inner-decoded.
\item $K\dfn\sum_{m\in[M]}K_{m}\in[N+1]$ is the total number of molecules
which have been erroneously inner-decoded. 
\item $V_{m}\in[K+1]$ is the number of output molecules $y^{L}$ that originated
from reading a molecule $x_{m'}^{L}(j)$ for some $m'\in[M]\backslash\{m\}$,
and that have been erroneously inner-decoded to have index $m$. Note
that $\sum_{m\in[M]}V_{m}\leq K$ holds.
\end{itemize}
With these definitions, the event in which the molecule $x_{m}^{L}$
was not decoded correctly in the second stage of the operation of
the decoder is included in a union of the following events:
\begin{enumerate}
\item $S_{m}<T_{\tau}$, that is, the molecule have not been sampled enough
times in the sampling stage. 
\item $S_{m}\geq T_{\tau}$ yet $S_{m}-K_{m}<T_{\tau}$, that is, the molecule
has been sampled enough times in the sampling stage step, but $K_{m}$
sequencing errors have caused the number of appearances of $x_{m}^{L}(j)$
to drop below the threshold $T_{\tau}$. 
\item $V_{m}\geq T_{\tau}$, that is, there are more than $T_{\tau}$ molecules
with index $m$, which are not the correct molecule $x_{m}^{L}(j)$.
\end{enumerate}
In correspondence to the three types of events, we define the following
sets for the analysis of multiple erasures/undetected-errors:
\begin{align}
{\cal M}_{\text{sam}} & \dfn\left\{ m\in[M]\colon S_{m}<T_{\tau}\right\} ,\label{eq: index sets 1}\\
{\cal M}_{\text{cbt}} & \dfn\left\{ m\in[M]\colon S_{m}\geq T_{\tau},\;S_{m}-K_{m}<T_{\tau}\right\} ,\label{eq: index sets 2}\\
{\cal M}_{\text{eat}} & \dfn\left\{ m\in[M]\colon V_{m}\geq T_{\tau}\right\} ,\label{eq: index sets 3}
\end{align}
where the subscripts are mnemonics to ``sampling'', ``correct-below-threshold'',
``erroneous-above-threshold''. We next make two remarks regarding
the definition of the above events:
\begin{enumerate}
\item The threshold rule in \eqref{eq: decoder threshold definition} requires
that the number of occurrences of the chosen molecule $b^{L}\in{\cal B}_{m}^{(L)}$
in $z^{LN}$ is large by at least $1$ over the maximal number of
occurrences of other competing molecules in ${\cal B}_{m}^{(L)}$.
Therefore, given a total of $K$ sequencing errors (out of $N$ molecules),
there could be at most $K$ molecules which were sampled a sufficient
number of times, i.e., $S_{m}\geq T_{\tau}$, yet sequencing errors
have caused the number of appearances of this molecule after the first
stage of decoding to drop below the threshold, i.e., the event $S_{m}-K_{m}<T_{\tau}$.
This bound seems rather crude, but will be shown to be effective in
what follows.
\item On the face of it, the third error event $V_{m}\geq T_{\tau}$ can
lead to a crude upper bound. In typical situations, the $V_{m}$ molecules
which are erroneously mapped to index $m$ due to sequencing errors
are not likely to be the exact same molecule in ${\cal B}_{m}^{(L)}$.
However, a more precise analysis of this event would require making
assumptions on the structure of the sub-codes $\{{\cal B}_{m}^{(L)}\}$,
which we avoid here altogether. In other words, we take here a worst-case
approach, and assume the most unfavorable situation in which the erroneous
$V_{m}$ molecules from ${\cal B}_{m}^{(L)}$ are the exact same molecule,
and whenever their number of appearances is larger than $T_{\tau}$,
an erasure occurs. 
\end{enumerate}
Finally, we define the set of erased molecules and the set of molecules
with undetected errors as
\begin{align}
{\cal M}_{\mathsf{e}} & \dfn\left\{ m\in[M]\colon\hat{x}_{m}^{L}=\mathsf{e}\right\} ,\label{eq: erasure set}\\
{\cal M}_{\mathsf{u}} & \dfn\left\{ m\in[M]\colon\hat{x}_{m}^{L}\neq\mathsf{e},\;\hat{x}_{m}^{L}\neq x_{m}^{L}(j)\right\} ,\label{eq: undetected error set}
\end{align}
assuming that the $j$th codeword was stored. Our goal in this section
is to bound the probability that $|{\cal M}_{\mathsf{e}}|$ is larger
than $\theta M$ for some $\theta\in[0,1]$, and the same probability
for $|{\cal M}_{\mathsf{u}}|$. To this end, we begin by deriving
relations between $|{\cal M}_{\text{sam}}|,|{\cal M}_{\text{cbt}}|$
and $|{\cal M}_{\text{eat}}|$ to the erasure and undetected error
sets. 
\begin{lem}
\label{lem:Bounds on the size of erasure and undetected sets}Let
$\tau\in(0,\frac{1}{2})$ be given. Then, 
\begin{equation}
|{\cal M}_{\mathsf{e}}|\leq|{\cal M}_{\text{sam}}|+\left(1+\frac{1}{T_{\tau}}\right)K\label{eq: bound on the number of erased molecules}
\end{equation}
and
\begin{equation}
|{\cal M}_{\mathsf{u}}|\leq\frac{K}{T_{\tau}}.\label{eq: bound on the number of undetected error molecules}
\end{equation}
\end{lem}
\begin{IEEEproof}
The bound \eqref{eq: bound on the number of erased molecules} is
proved by the following chain of inequalities:
\begin{align}
|{\cal M}_{\mathsf{e}}| & \leq|{\cal M}_{\text{sam}}\cup{\cal M}_{\text{cbt}}\cup{\cal M}_{\text{eat}}|\\
 & \leq|{\cal M}_{\text{sam}}|+|{\cal M}_{\text{cbt}}|+|{\cal M}_{\text{eat}}|\\
 & \trre[\leq,a]|{\cal M}_{\text{sam}}|+K+\frac{K}{T_{\tau}},
\end{align}
where the inequality follows since both $|{\cal M}_{\text{cbt}}|\leq K$
and $|{\cal M}_{\text{eat}}|\leq\frac{K}{T_{\tau}}$ must hold at
worst case (as discussed informally above). The bound \eqref{eq: bound on the number of undetected error molecules}
follows immediately from the threshold definition in \eqref{eq: threshold definition}
as 
\[
|{\cal M}_{\mathsf{u}}|\leq|{\cal M}_{\text{eat}}|\leq\frac{K}{T_{\tau}}.
\]
\end{IEEEproof}
The next pair of lemmas is devoted to analyzing the probability that
the cardinality of one of the index sets defined in \ref{eq: index sets 1},
\ref{eq: index sets 2} and \eqref{eq: index sets 3}, is larger than
some threshold. We begin with the cardinality of ${\cal M}_{\text{sam}}$. 
\begin{lem}
\label{lem: amplification large deviations}Let $\tau\in(0,\frac{1}{2})$
be given, and let $x^{LM}(j)$ be a codeword from a coded-index codebook.
Let $\tilde{S}\sim\text{\emph{Pois}}(N/M)$ and
\[
\varphi_{\tau}\dfn-\frac{1}{N/M}\log\P\left[\tilde{S}\leq T_{\tau}\right].
\]
Let ${\cal M}_{\text{sam}}$ be as in \eqref{eq: index sets 1}. If
$N/M=\Theta(1)$ then 
\[
\P\left[|{\cal M}_{\text{sam}}|\geq\sigma M\mid x^{LM}(j)\right]\leq2(1+o(1)\cdot\exp\left[-M\cdot d_{b}\left(\sigma\,\middle\vert\middle\vert\,e^{-\varphi_{\tau}\frac{N}{M}}\right)\right]
\]
for $\sigma\in(e^{-\varphi_{\tau}\frac{N}{M}},1]$. If $N/M=\omega(1)$
then 
\[
\P\left[|{\cal M}_{\text{sam}}|\geq\sigma M\mid x^{LM}(j)\right]\leq4e^{-\sigma\tau N\cdot[1+o(1)]}.
\]
for $\sigma\in(e^{-\tau\frac{N}{M}},1]$.
\end{lem}
\begin{IEEEproof}
The molecule duplicate vector $S^{M}$ follows a multinomial distribution,
and so its components $\{S_{m}\}_{m\in[M]}$ are dependent random
variables. To facilitate the analysis, we use the following well-known
fact (e.g., \cite[Thm. 5.6]{mitzenmacher2017probability}):
\begin{fact}[Poissonization of the multinomial distribution]
Let $\tilde{N}\sim\text{\emph{Pois}}(\lambda)$, and let $\tilde{S}^{M}$
be a random vector such that $\tilde{S}^{M}\sim\text{\emph{Multinomial}}(\tilde{N},(p_{0},\ldots p_{M-1}))$
conditioned on $\tilde{N}$, where $\sum_{m\in[M]}p_{m}=1$ and $p_{m}>0$.
Then, $\{\tilde{S}_{m}\}_{m\in[M]}$ are statistically independent
and $\tilde{S}_{m}\sim\text{\emph{Pois}}(p_{m}\lambda)$ (unconditioned
on $\tilde{N}$). 
\end{fact}
Returning to the DNA storage model, let $\tilde{N}\sim\text{Pois}(N)$
and $\tilde{S}^{M}=(\tilde{S}_{0}\ldots,\tilde{S}_{M-1})\sim\text{Multinomial}(\tilde{N},(\frac{1}{M},\ldots,\frac{1}{M}))$
conditioned on $\tilde{N}$. By the above fact, $\{\tilde{S}_{m}\}_{m\in[M]}$
are i.i.d. and $\tilde{S}_{m}\sim\text{Pois}(\frac{N}{M})$. Define
the event $\xi_{m}\dfn\I\{S_{m}<T_{\tau}\}$, and similarly the event
$\tilde{\xi}_{m}\dfn\I\{\tilde{S}_{m}<T_{\tau}\}$, where $T_{\tau}=\frac{N}{M}(1-\sqrt{2\tau})$
is the threshold for some $\tau\in(0,\frac{1}{2})$. Thus, clearly,
$T_{\tau}\leq\E[\tilde{S}_{m}]=\frac{N}{M}$, and so the expected
number of occurrences of the $m$th molecule at the output exceeds
the threshold for noiseless sequencing. 

We begin with the $N/M=\Theta(1)$ case. In this regime, $\P[\tilde{S}_{m}\leq T_{\tau}]$
can be computed directly, and it specifically holds by the definition
of $\varphi_{\tau}$ that $\P[\tilde{S}_{m}\leq T_{\tau}]=e^{-\varphi_{\tau}\frac{N}{M}}$.
Next, we derive an inequality which will be used to bound the probability
that $\sum_{m\in[M]}\xi_{m}\geq\sigma M$. Recall that the sum of
two independent multinomial distributions with the same probability
parameters $(p_{0},\ldots p_{M-1})$ and $N_{1}$ (resp. $N_{1}'$)
trials is distributed as $\text{Multinomial}(N_{1}+N_{1}',(p_{0},\ldots p_{M-1}))$.
Thus, if $N_{1}\leq N_{2}=N_{1}+N_{1}'$ then 
\[
\P\left[\sum_{m\in[M]}\I\{\tilde{S}_{m}<T_{\tau}\}\geq\sigma M\,\middle\vert\,\tilde{N}=N_{1}\right]\geq\P\left[\sum_{m\in[M]}\I\{\tilde{S}_{m}<T_{\tau}\}\geq\sigma M\,\middle\vert\,\tilde{N}=N_{2}\right]
\]
holds. A simple application of the law of total expectation  then
implies that 
\begin{equation}
\P\left[\sum_{m\in[M]}\I\{\tilde{S}_{m}<T_{\tau}\}\geq\sigma M\,\middle\vert\,\tilde{N}=N\right]\leq\P\left[\sum_{m\in[M]}\I\{\tilde{S}_{m}<T_{\tau}\}\geq\sigma M\,\middle\vert\,\tilde{N}\leq N\right].\label{eq: monotinicity of large deviations with the number of trials}
\end{equation}
Hence, the required probability is bounded as (see also \cite[Exercise 5.14]{mitzenmacher2017probability})
\begin{align}
\P\left[\sum_{m\in[M]}\xi_{m}\geq\sigma M\right] & =\P\left[\sum_{m\in[M]}\I\{S_{m}<T_{\tau}\}\geq\sigma M\right]\\
 & =\P\left[\sum_{m\in[M]}\I\{\tilde{S}_{m}<T_{\tau}\}\geq\sigma M\,\middle\vert\,\tilde{N}=N\right]\\
 & \trre[\leq,a]\P\left[\sum_{m\in[M]}\I\{\tilde{S}_{m}<T_{\tau}\}\geq\sigma M\,\middle\vert\,\tilde{N}\leq N\right]\\
 & =\frac{\P\left[\sum_{m\in[M]}\I\{\tilde{S}_{m}<T_{\tau}\}\geq\sigma M,\;\tilde{N}\leq N\right]}{\P\left[\tilde{N}\le N\right]}\\
 & \leq\frac{\P\left[\sum_{m\in[M]}\I\{\tilde{S}_{m}<T_{\tau}\}\geq\sigma M\right]}{\P\left[\tilde{N}\le N\right]}\\
 & \trre[\leq,b]2\cdot(1+o(1))\cdot\P\left[\sum_{m\in[M]}\I\{\tilde{S}_{m}<T_{\tau}\}\geq\sigma M\right]\\
 & =2\cdot(1+o(1))\cdot\P\left[\sum_{m\in[M]}\tilde{\xi}_{m}\geq\sigma M\right]\\
 & \trre[\leq,c]2\cdot(1+o(1))\cdot\exp\left[-M\cdot d_{b}\left(\sigma||\E[\tilde{\xi}_{m}]\right)\right]\\
 & \leq2\cdot(1+o(1))\cdot\exp\left[-M\cdot d_{b}\left(\sigma||e^{-\varphi_{\tau}N/M}\right)\right],\label{eq: Poissonization tail bound}
\end{align}
where $(a)$ follows from \eqref{eq: monotinicity of large deviations with the number of trials},
$(b)$ follows from the central limit theorem for Poisson random variables
which states that $\frac{\tilde{N}-N}{\sqrt{N}}\to{\cal N}(0,1)$
in distribution as $N\to\infty$ and so $\P[\tilde{N}\leq N]\to\frac{1}{2}$,
and $(c)$ follows since the random variables $\tilde{\xi}_{m}=\I\{\tilde{s}_{m}\leq T\}$,
$m\in[M]$ are i.i.d., and so Chernoff's bound for the binomial distribution
implies this bound for any $\sigma\in(e^{-\varphi_{\tau}N/M},1]$.
The claimed bound then follows for $N/M=\Theta(1)$. 

Next, for $N/M=\omega(1)$ we utilize Chernoff's bound for Poisson
random variables. Specifically, if $A\sim\text{Pois}(\lambda)$ then
for any $a<\lambda$ it holds that \cite[Thm. 5.4]{mitzenmacher2017probability}
\[
\P[A\leq a]\leq\left(\frac{e\lambda}{a}\right)^{a}e^{-\lambda}.
\]
Then, setting $a=c\lambda$ for $c\in[0,1)$ we obtain 
\begin{align}
\P[A\leq c\lambda] & \leq\left(\frac{e}{c}\right)^{a}e^{-\lambda}\\
 & =e^{c\lambda\log(e/c)-\lambda}\\
 & =e^{-\lambda\left[c\log c-c+1\right]}\\
 & \trre[\leq,a]\exp\left[-\lambda\left(c\left(c-1-\frac{1}{2}(c-1)^{2}\right)-c+1\right)\right]\\
 & =\exp\left[-\lambda\left((c-1)^{2}-\frac{c}{2}(c-1)^{2}\right)\right]\\
 & \leq\exp\left[-\frac{\lambda}{2}(c-1)^{2}\right],\label{eq: Poisson - Gaussian tails}
\end{align}
where $(a)$ follows from $\log c\geq c-1-\frac{1}{2}(c-1)^{2}$ for
$c\in[0,1]$. Then, using \eqref{eq: Poisson - Gaussian tails}, we
obtain 
\begin{align}
\E[\tilde{\xi}_{m}] & =\P[\tilde{S}_{m}\leq T_{\tau}]\\
 & \leq\exp\left[-\frac{N}{2M}\left(\frac{T_{\tau}M}{N}-1\right)^{2}\right]\\
 & =\exp\left[-\frac{\tau N}{M}\right],
\end{align}
that is, the bound $\varphi_{\tau}\geq\tau$ holds. We then approximate
\begin{equation}
\exp\left[-M\cdot d_{b}\left(\sigma||e^{-\varphi_{\tau}N/M}\right)\right]\leq\exp\left[-M\cdot d_{b}\left(\sigma||e^{-\tau N/M}\right)\right]=\exp\left[-\sigma\tau N\cdot[1+o(1)]\right],\label{eq: approximiation of expoeent for large N over M}
\end{equation}
using the asymptotic expansion of the binary KL divergence $d_{b}(a||b)=-[1+o(1)]\cdot a\log b$
(see Proposition \ref{prop: Asymptotic expansions of the binary KL divergence}
in Appendix \ref{sec:Asymptotic-Expansions-of}). The result then
follows by an analysis similar to the $N/M=\Theta(1)$ case above
until \eqref{eq: Poissonization tail bound}, followed by an application
of the bound \eqref{eq: approximiation of expoeent for large N over M}.
\end{IEEEproof}
In the next lemma, we bound the large deviations of the total number
of sequencing errors $K$. To this end, we first discuss the statistical
dependency between sampling events and sequencing errors events. In
principle, the random behavior of the sequencing channel is independent
of the sampling operation. However, since the error probability of
the inner-code ${\cal B}^{(L)}$ may be different when conditioned
on different codewords, the total number of sequencing errors may
depend on the sampling event.\footnote{For example, consider the case in which the sub-code ${\cal B}_{1}^{(L)}$
has a low average error probability compared to the other $M-1$ sub
codes $\{{\cal B}_{m}^{(L)}\}_{m\in[M]\backslash\{1\}}$. In this
case, the expected value of the number of sequencing errors conditioned
the sampling event $S_{1}=M$ is lower than the unconditional expected
value. } Following our general approach in this paper, we avoid assuming any
structure on the inner code, and so we take a worst-case approach.
Let $\overline{b}^{L}\in{\cal B}^{(L)}$ be the inner-code codeword
which achieves the maximal error probability over all possible codewords
of the code ${\cal B}^{(L)}$, and let $\overline{Y}^{(L)}\in{\cal Y}^{(L)}$
be a random sequenced output given that $\overline{b}^{L}$ was input
to $W^{(L)}$. Then, 
\[
\pe_{b}({\cal B}^{(L)})=W^{(L)}\left[\mathsf{D}_{b}(\overline{Y}^{(L)})\neq\overline{b}^{L}\mid\overline{b}^{L}\right]=\max_{b^{L}\in{\cal B}^{(L)}}W^{(L)}\left[\mathsf{D}_{b}(\overline{Y}^{(L)})\neq b^{L}\mid b^{L}\right].
\]
We define by $\tilde{K}$ the total number of molecules which were
erroneously sequenced, conditioned on the event that the only sampled
molecule is $\overline{b}^{L}$. Then, clearly $\tilde{K}\sim\text{Binomial}(N,\pe_{b}({\cal B}^{(L)}))$,
and $\tilde{K}$ is independent of the sampling event. 
\begin{lem}
\label{lem: total seqeuncing errors large deviations}Let $K$ be
the total number of erroneously sequenced molecules out of the $N$
sampled molecules. Let ${\cal U}\subset[M]^{N}$ be an arbitrary sampling
event, and assume that $\pe_{b}({\cal B}^{(L)})=e^{-c\cdot L^{\zeta}}$.
Then, for any $\kappa\in(0,1]$

\[
\P\left[K\geq\kappa N\mid{\cal U}\right]\leq\exp\left(-c(1+o(1))\cdot\kappa NL^{\zeta}\right).
\]
\end{lem}
\begin{IEEEproof}
Assume that an arbitrary codeword $x^{LM}$ has been stored. Let $E_{n}=\I[\mathsf{D}_{b}(Y_{n}^{(L)})\neq x_{U_{n}}^{L}]$
denote the indicator of the event that a sequencing error has occurred
in the first stage of the decoder, for the $n$th sampled molecule,
$n\in[N]$. Then, conditioned on any sampling event ${\cal U}$,
\begin{align}
\P\left[K\geq\kappa M\mid{\cal U}\right] & \leq\P\left[\sum_{n\in[N]}E_{n}\ge\kappa M\,\middle\vert\,{\cal U}\right]\\
 & \trre[\leq,*]\P\left[\tilde{K}\ge\kappa M\mid{\cal U}\right]\\
 & =\P\left[\tilde{K}\ge\kappa M\right],\label{eq: justification of independence of sampling and sequencing errors}
\end{align}
where $(*)$ follows from the following consideration: Let $E^{N}\in\{0,1\}^{N}$
be a sequence of independent Bernoulli trials so that $\P[E_{n}=1]=p_{n}$.
Let $p_{\text{max}}\dfn\max_{n\in[N]}p_{n}$, and let $A\sim\text{Binomial}(N,p_{\text{max}})$.
Let $\tilde{E}^{N}\in\{0,1\}^{N}$ be another sequence of independent
Bernoulli trials, statistically independent of $E^{N}$, so that $\P[E_{n}\vee\tilde{E}_{n}=1]=p_{\max}$
for all $n\in[N]$ (concretely, $\P[\tilde{E}_{n}=1]=\tilde{p}_{n}=(p_{\text{max}}-p_{n})/(1-p_{n})$).
Then, 
\[
\P\left[\sum_{n\in[N]}E_{n}\ge t\right]\leq\P\left[\sum_{n\in[N]}(E_{n}\vee\tilde{E}_{n})\ge t\right]=\P\left[A\ge t\right].
\]
Given \eqref{eq: justification of independence of sampling and sequencing errors},
we may next analyze the large-deviations of $\tilde{K}$ in lieu of
that of $K$. For any fixed $\kappa\in(0,1]$, the expected number
of sequencing errors is 
\[
\E[\tilde{K}]=Ne^{-c\cdot L^{\zeta}}=o(\kappa N).
\]
Thus, the event $\tilde{K}\geq\kappa N$ is a large-deviations event.
Since $\tilde{K}\sim\text{Binomial}(N,e^{-c\cdot L^{\zeta}})$, Chernoff's
bound implies that 
\[
\P\left[\tilde{K}\geq\kappa N\right]\leq\exp\left(-N\cdot d_{b}\left(\kappa\,\middle\vert\middle\vert\,e^{-c\cdot L^{\zeta}}\right)\right),
\]
and the result follows from the asymptotic expansion of the binary
KL divergence $d_{b}(a||b)=-[1+o(1)]\cdot a\log b$ (see Proposition
\ref{prop: Asymptotic expansions of the binary KL divergence} in
Appendix \ref{sec:Asymptotic-Expansions-of}).
\end{IEEEproof}
We next utilize Lemmas \ref{lem: amplification large deviations},
and \ref{lem: total seqeuncing errors large deviations} to obtain
the large-deviations behavior of the cardinality of erasure and undetected
errors sets. As will be apparent, the dominating event is $\P[|{\cal M}_{\text{sam}}|\geq\sigma M]$
-- to wit, the probability that the molecules have not been amplified
enough times -- which is on the exponential order of $N$, compared
to the probability evaluated in Lemma \ref{lem: total seqeuncing errors large deviations}
which is on the exponential order of $LN=N\beta\log M$. Therefore,
as also discussed in the introduction, for the coded-index based codebooks
and the type of decoders studied here, the effect of sequencing errors
is much less profound compared to erasures. 
\begin{lem}
\label{lem: erausre large deviations with high threshold}Let $\tau\in(0,\frac{1}{2})$
be given. Consider a decoder $\mathsf{D}$ for a coded-index based
codebook. For the erasure set ${\cal M}_{\mathsf{e}}$:
\begin{itemize}
\item If $N/M=\Theta(1)$
\[
-\log\P\left[|{\cal M}_{\mathsf{e}}|\geq\theta M\right]\geq M\cdot d_{b}\left(\theta||e^{-\varphi_{\tau}\frac{N}{M}}\right)+o(M)
\]
for all $\theta\in(e^{-\varphi_{\tau}\frac{N}{M}},1]$.
\item If $N/M=\omega(1)$ then 
\[
-\log\P\left[|{\cal M}_{\mathsf{e}}|\geq\theta M\right]\geq\theta\tau N\cdot[1+o(1)]
\]
for all $\theta\in(e^{-\tau\frac{N}{M}},1]$.
\end{itemize}
Furthermore, for the undetected error set ${\cal M}_{\mathsf{u}}$:\textbf{
\[
-\log\P\left[|{\cal M}_{\mathsf{u}}|\geq\theta M\right]\geq c(1+o(1))\cdot(1-\sqrt{2\tau})\theta NL^{\zeta}.
\]
}

\end{lem}
\begin{IEEEproof}
For any $\theta,\kappa\in\frac{1}{M}\cdot[M+1]$ 
\begin{align}
 & \P\left[|{\cal M}_{\mathsf{e}}|\geq\theta M\right]\nonumber \\
 & \trre[\leq,a]\P\left[|{\cal M}_{\text{sam}}|+\left(1+\frac{1}{T_{\tau}}\right)K\geq\theta M\right]\\
 & \trre[\leq,b]\sum_{\sigma,\kappa\in\frac{1}{M}\cdot[M+1]\colon\sigma+\kappa\geq\theta}\P\left[|{\cal M}_{\text{sam}}|\geq\sigma M\right]\cdot\P\left[\left(1+\frac{1}{T_{\tau}}\right)K\geq\kappa M\,\middle\vert\,|{\cal M}_{\text{sam}}|=\sigma M\right]\\
 & \trre[\leq,c]\sum_{\sigma,\kappa\in\frac{1}{M}\cdot[M+1]\colon\sigma+\kappa\geq\theta}\P\left[|{\cal M}_{\text{sam}}|\geq\sigma M\right]\cdot\exp\left(-c(1+o(1))\cdot\frac{\kappa\left(1\wedge T_{\tau}\right)}{2}ML^{\zeta}\right)\label{eq: number of erasures with sequencing errors}
\end{align}
where $(a)$ follows from Lemma \ref{lem:Bounds on the size of erasure and undetected sets},
$(b)$ follows from the union bound, $(c)$ follows from Lemma \ref{lem: total seqeuncing errors large deviations}
by
\begin{align}
 & \P\left[\left(1+\frac{1}{T_{\tau}}\right)K\geq\kappa M\,\middle\vert\,|{\cal M}_{\text{sam}}|=\sigma M\right]\nonumber \\
 & =\P\left[K\geq\frac{\kappa}{\frac{N}{M}\left(1+\frac{1}{T_{\tau}}\right)}N\,\middle\vert\,|{\cal M}_{\text{sam}}|=\sigma M\right]\\
 & \leq\exp\left(-c(1+o(1))\cdot\frac{\kappa}{\frac{N}{M}\left(1+\frac{1}{T_{\tau}}\right)}NL^{\zeta}\right)\\
 & \leq\exp\left(-c(1+o(1))\cdot\frac{\kappa\left(1\wedge T_{\tau}\right)}{2}ML^{\zeta}\right)\\
 & \leq\exp\left(-c(1+o(1))\cdot\frac{\kappa\left(1\wedge T_{\tau}\right)}{2}ML^{\zeta}\right),
\end{align}
where $c>0$ is the constant for which $\pe_{b}({\cal B}^{(L)})\leq e^{-c\cdot L^{\zeta}}$.
We continue to bound the probability of interest for the $N/M=\Theta(1)$
case. Using Lemma \ref{lem: amplification large deviations} in \eqref{eq: number of erasures with sequencing errors}
\[
\P\left[|{\cal M}_{\mathsf{e}}|\geq\theta M\right]\leq2(1+o(1))\cdot M^{2}\times\max_{\sigma,\kappa\in[0,1]\colon\sigma+\kappa\geq\theta}\exp\left[-M\cdot\left(d_{b}\left(\sigma\,\middle\vert\middle\vert\,e^{-\varphi_{\tau}\frac{N}{M}}\right)+\frac{\kappa\left(1\wedge T_{\tau}\right)}{2}ML^{\zeta}\right)\right]
\]
and so 
\[
-\frac{1}{M}\log\P\left[|{\cal M}_{\mathsf{e}}|\geq\theta M\right]\geq\min_{\sigma,\kappa\in[0,1]\colon\sigma+\kappa\geq\theta}d_{b}\left(\sigma\,\middle\vert\middle\vert\,e^{-\varphi_{\tau}\frac{N}{M}}\right)+\frac{\kappa\left(1\wedge T_{\tau}\right)}{2}L^{\zeta}-O\left(\frac{\log M}{M}\right)
\]
for all $\sigma\in(e^{-\varphi_{\tau}\frac{N}{M}},1]$. In the last
minimization problem, any choice of $\kappa>0$ will cause the exponent
to be $\Theta(L^{\zeta})=\Theta(\log^{\zeta}M)$ and thus diverge
as $M\to\infty$. Thus, the minimum is obtained for $\kappa=o(1)$
and if $\theta\geq e^{-\varphi_{\tau}\frac{N}{M}}$ the minimum is
obtained at $\sigma=\theta-o(1)$. The claimed result then follows
since 
\[
d_{b}\left(\sigma\,\middle\vert\middle\vert\,e^{-\varphi_{\tau}\frac{N}{M}}\right)=d_{b}\left(\theta-o(1)\,\middle\vert\middle\vert\,e^{-\varphi_{\tau}\frac{N}{M}}\right)=d_{b}\left(\theta\,\middle\vert\middle\vert\,e^{-\varphi_{\tau}\frac{N}{M}}\right)-o(1)
\]
for any fixed $\tau$, and since $N/M=\Theta(1)$ was assumed. If
$\theta\leq e^{-\varphi_{\tau}\frac{N}{M}}$ then the exponent is
$O(\frac{\log M}{M})$. The analysis for the $N/M=\omega(1)$ case
is analogous and thus omitted. 

Next, the probability that the cardinality of the undetected error
set exceeds a certain threshold is bounded as: 
\begin{align}
\P\left[|{\cal M}_{\mathsf{u}}|\geq\theta M\right] & \trre[\leq,a]\P\left[K\geq(1-\sqrt{2\tau})\theta N\right]\\
 & \trre[\leq,b]\exp\left[-c(1+o(1))\cdot(1-\sqrt{2\tau})\theta NL^{\zeta}\right]
\end{align}
where $(a)$ is using Lemma \ref{lem:Bounds on the size of erasure and undetected sets},
and $(b)$ is by Lemma \ref{lem: total seqeuncing errors large deviations}.
\end{IEEEproof}

\subsection{Random Coding Analysis \label{subsec:Random-Coding-Analysis}}

After analyzing the effect of the channel on the number of erasures
and undetected error events at the decoder (the probability that the
number of erasures or undetected errors exceeds $\theta M$), we turn
to the analysis of the average error probability of the random coding
ensemble. We let $\overline{\pe}({\cal C}_{M},\mathsf{D}_{M})$ denote
the ensemble average over the choice of random codebook ${\cal C}_{M}$
from the ensemble of Definition \ref{def:RCE}. The random coding
bound is as follows:
\begin{prop}[Random coding bound]
\label{prop: index molecule random coding bound} Let an inner code
be given by $\{{\cal B}_{m}^{(L)}\}\subset{\cal X}^{L}$ for $m\in[M]$,
and let $\mathsf{D}_{b}$ be a decoder which satisfy the assumption
on the inner code at rate $R_{b}$. The random coding exponent over
the random ensemble from Definition \ref{def:RCE}, with the decoder
described in Sec. \ref{subsec:Formulation-of-the} is bounded as follows:
\begin{itemize}
\item If $N/M=\Theta(1)$ then 
\[
-\liminf_{M\to\infty}\frac{1}{M}\log\overline{\pe}({\cal C}_{M},\mathsf{D}_{M})\geq d_{b}\left(1-\frac{R}{R_{b}-1/\beta}\,\middle\vert\middle\vert\,e^{-\frac{N}{M}}\right)
\]
for any $R<(R_{b}-1/\beta)(1-e^{-\frac{N}{M}})$. 
\item If $N/M=\omega(1)$ then 
\[
-\liminf_{N\to\infty}\frac{1}{N}\log\overline{\pe}({\cal C}_{M},\mathsf{D}_{M})\geq\begin{cases}
\frac{1}{2}\left[1-\frac{R}{R_{b}-1/\beta}\right], & \frac{N}{ML}\leq2[R_{b}-1/\beta]\\
\frac{ML}{N}\cdot\left[R_{b}-1/\beta-R\right], & \frac{N}{ML}>2[R_{b}-1/\beta]
\end{cases}
\]
for any $R<R_{b}-1/\beta$.
\end{itemize}
\end{prop}
In the rest of the section we prove Prop. \ref{prop: index molecule random coding bound}.
The next lemma bounds the probability that an erroneous codeword will
be decoded conditioned on a given number of channel erasures and undetected
errors. 
\begin{lem}
\label{lem: random coding pairwise error probability}Let ${\cal C}$
be drawn from the coded-index based random coding ensemble (Definition
\ref{def:RCE}). Let $X^{LM}(0)=x^{LM}(0)$ be arbitrary, and let
$\hat{X}^{LM}$ be the output of the decoder conditioned on the input
$x^{LM}(0)$. Then, for $\theta_{\mathsf{e}},\theta_{\mathsf{u}}\in\frac{1}{M}[M+1]$
such that $\theta_{\mathsf{e}}+\theta_{\mathsf{u}}\leq1$ and any
$j\in[|{\cal C}|]\backslash\{0\}$ it holds that 
\begin{multline}
-\frac{1}{M}\log\P\left[\rho(\hat{X}^{LM},X^{LM}(j))\leq\rho(\hat{X}^{LM},x^{LM}(0))\,\middle\vert\,|{\cal M}_{\mathsf{e}}|=\theta_{\mathsf{e}}M,\;|{\cal M}_{\mathsf{u}}|=\theta_{\mathsf{u}}M\right]\\
\geq(R_{b}\beta-1)(1-\theta_{\mathsf{e}}-\theta_{\mathsf{u}})\log M-1+\frac{\log M}{M}.
\end{multline}
\end{lem}
\begin{IEEEproof}
We may assume without loss of generality (w.l.o.g.) that ${\cal M}_{\mathsf{e}}=[\theta_{\mathsf{e}}M]$
and that ${\cal M}_{\mathsf{u}}=[(\theta_{\mathsf{e}}+\theta_{\mathsf{u}})M]\backslash[\theta_{\mathsf{e}}M]$.
Let $R_{b}=\frac{1}{L}\log|{\cal B}^{(L)}|$ be the rate of the inner
code. By the definition of the random ensemble, the total number of
possibilities to choose the codeword $X^{LM}(j)$ is $\left(e^{R_{b}L}/M\right)^{M}=M^{M(R_{b}\beta-1)}$
where $R_{b}\leq\log|{\cal X}|$. Let $\hat{X}^{LM}$ be an arbitrary
decoder output such that the event $\{|{\cal M}_{\mathsf{e}}|=\theta_{\mathsf{e}}M,\;|{\cal M}_{\mathsf{u}}|=\theta_{\mathsf{u}}M\}$
holds. We next upper bound the number of possible codewords that result
a distance $\rho(\hat{X}^{LM},X^{LM}(j))$ that is no larger than
$\rho(\hat{X}^{LM},x^{LM}(0))=\theta_{\mathsf{u}}M$ as follows: For
the indices $m\in{\cal M}_{\mathsf{e}}$ the choice of $X_{m}(j)$
may be arbitrary since it does not affect the distance $\rho(\hat{X}^{LM},X^{LM}(j))$.
The number of possibilities to choose molecules for these indices
is$\left(e^{R_{b}L}/M\right)^{\theta_{\mathsf{e}}M}=M^{\theta_{\mathsf{e}}M(R_{b}\beta-1)}$.
Then, the codeword $X^{LM}(j)$ will cause an error if the $\rho$-distance
at the remaining set of indices $[M]\backslash{\cal M}_{\mathsf{e}}$
is less or equal to $\theta_{\mathsf{u}}M$. The number of possibilities
for this choice is 
\[
\sum_{r=0}^{\theta_{\mathsf{u}}M}{M(1-\theta_{\mathsf{e}}) \choose r}\cdot\left(\frac{e^{R_{b}L}}{M}-1\right)^{r},
\]
where in the summation above $r$ is the resulting distance $\rho(\hat{X}^{LM},X^{LM}(j))$,
${M(1-\theta_{\mathsf{e}}) \choose r}$ is the number of possibilities
to choose a subset of cardinality $r$ from the set of indices $\{m\in[M]\backslash{\cal M}_{\mathsf{e}}:X_{m}^{L}(j)\neq\hat{X}^{L}\}$,
and $(e^{R_{b}L}/M-1)^{r}$ is the number of ways to choose molecules
for these indices. Since the choice of codeword $X^{LM}(j)$ is uniform
over the set of all possibilities, it holds that
\begin{align}
 & \P\left[\rho(\hat{X}^{LM},X^{LM}(j))\leq\rho(\hat{X}^{LM},X^{LM}(0))\,\middle\vert\,{\cal M}_{\mathsf{e}},{\cal M}_{\mathsf{u}}\right]\nonumber \\
 & \leq\frac{M^{\theta_{\mathsf{e}}M(R_{b}\beta-1)}\cdot\sum_{r=0}^{\theta_{\mathsf{u}}M}{M(1-\theta_{\mathsf{e}}) \choose r}\cdot\left(\frac{e^{R_{b}L}}{M}-1\right)^{r}}{M^{M(R_{b}\beta-1)}}\\
 & \trre[\leq,a]\sum_{r=0}^{\theta_{\mathsf{u}}M}{M(1-\theta_{\mathsf{e}}) \choose r}\cdot M^{(R_{b}\beta-1)(\theta_{\mathsf{e}}+\frac{r}{M}-1)M}\\
 & \trre[\leq,b]\theta_{\mathsf{u}}M\cdot{M(1-\theta_{\mathsf{e}}) \choose \theta_{\mathsf{u}}M\wedge\frac{M(1-\theta_{\mathsf{e}})}{2}}M^{(R_{b}\beta-1)(\theta_{\mathsf{e}}+\theta_{\mathsf{u}}-1)M}\\
 & \trre[\leq,c]Me^{M(1-\theta_{\mathsf{e}})}M^{(R_{b}\beta-1)(\theta_{\mathsf{e}}+\theta_{\mathsf{u}}-1)M}\\
 & \leq Me^{M}M^{(R_{b}\beta-1)(\theta_{\mathsf{e}}+\theta_{\mathsf{u}}-1)M},
\end{align}
where $(a)$ follows from $\frac{e^{R_{b}L}}{M}-1\leq\frac{e^{R_{b}L}}{M}=\frac{e^{R_{b}\beta\log M}}{M}=M^{(R_{b}\beta-1)}$,
$(b)$ follows since the binomial coefficient ${n \choose k}$ is
monotonic increasing in $k$ for $k\leq n/2$, $(c)$ follows from
the bound ${n \choose k}\leq e^{nh_{b}(k/n)}\leq e^{n}$. The claim
of the lemma follows by rearranging the terms and noting that the
bound depends on ${\cal M}_{\mathsf{e}},{\cal M}_{\mathsf{u}}$ only
via their respective cardinality $\theta_{\mathsf{e}},\theta_{\mathsf{u}}$.
\end{IEEEproof}
We next prove the random coding bound of Proposition \ref{prop: index molecule random coding bound}:
\begin{IEEEproof}[Proof of Proposition \ref{prop: index molecule random coding bound}]
By symmetry of the random coding ensemble w.r.t. the choice of codeword,
we may assume w.l.o.g. that $X^{LM}(0)=x^{LM}(0)$ was stored, and
condition on this event. We further condition that $\hat{X}^{LM}=\hat{x}^{LM}$
was obtained after the first two stages of decoding, and that $|{\cal M}_{\mathsf{e}}|=\theta_{\mathsf{e}}M$
and $|{\cal M}_{\mathsf{u}}|=\theta_{\mathsf{u}}M$ for some $\theta_{\mathsf{e}},\theta_{\mathsf{u}}\in[0,1]$.
By the clipped union bound and symmetry, the conditional average error
probability is 
\begin{align}
 & \P\left[\text{error}\mid X^{LM}(0)=x^{LM}(0),\;\hat{X}^{LM}=\hat{x}^{LM}\right]\nonumber \\
 & \leq1\wedge\sum_{j=2}^{e^{MLR}}\P\left[\rho(\hat{x}^{LM},X^{LM}(j))\leq\rho(\hat{x}^{LM},x^{LM}(0))\right]\\
 & \leq1\wedge e^{MLR}\cdot\P\left[\rho(\hat{x}^{LM},X^{LM}(j))\leq\rho(\hat{x}^{LM},x^{LM}(0))\right],\label{eq: conditional error probabilty exlcusive molecule random coding}
\end{align}
and so by Lemma \ref{lem: random coding pairwise error probability}
\begin{align}
 & -\log\P\left[\text{error}\mid X^{LM}(0)=x^{LM}(0),\;\hat{X}^{LM}=\hat{x}^{LM}\right]\nonumber \\
 & \geq M\cdot\left[-1+(R_{b}\beta-1)(1-\theta_{\mathsf{e}}-\theta_{\mathsf{u}})\log M-LR\right]_{+}-\log M\\
 & =M\cdot\left[(R_{b}\beta-1)(1-\theta_{\mathsf{e}}-\theta_{\mathsf{u}})\log M-L\left(R+\frac{1}{L}\right)\right]_{+}-\log M\\
 & \dfn M\cdot G_{R'}(\theta_{\mathsf{e}},\theta_{\mathsf{u}})-\log M,
\end{align}
where $G_{R'}(\theta_{\mathsf{e}},\theta_{\mathsf{u}})$ was implicitly
defined, and where $R'=R+\frac{1}{L}$. Clearly $R'\geq\frac{1}{L}$
must hold, however, we will continue the analysis without this constraint,
and then eventually take it into account. 

The bound above on the error probability depends on $\hat{x}^{LM},x^{LM}(0)$
only via $\theta_{\mathsf{e}},\theta_{\mathsf{u}}$, which satisfy
the constraint $\theta_{\mathsf{e}}+\theta_{\mathsf{u}}\leq1$. Let
$\overline{\pe}({\cal C}_{M},\mathsf{D}_{M})$ denote the ensemble
average error probability. Then, it is bounded as 
\begin{align}
 & \overline{\pe}({\cal C}_{M},\mathsf{D}_{M})\nonumber \\
 & \leq\sum_{\theta_{\mathsf{e}},\theta_{\mathsf{u}}\in\frac{1}{M}\cdot[M+1]}\P\left[|{\cal M}_{\mathsf{e}}|=\theta_{\mathsf{e}}M,|{\cal M}_{\mathsf{u}}|=\theta_{\mathsf{u}}M\right]\cdot e^{-M\cdot G_{R'}(\theta_{\mathsf{e}},\theta_{\mathsf{u}})+\log M}\\
 & \leq M^{2}\cdot\max_{\theta_{\mathsf{e}},\theta_{\mathsf{u}}\in[0,1]}\P\left[|{\cal M}_{\mathsf{e}}|\geq\theta_{\mathsf{e}}M,|{\cal M}_{\mathsf{u}}|\geq\theta_{\mathsf{u}}M\right]\cdot e^{-M\cdot G_{R'}(\theta_{\mathsf{e}},\theta_{\mathsf{u}})+\log M}.\label{eq: random coding bound with sequencing proof}
\end{align}
Let $\delta\in(0,1)$ be arbitrary. We next separate the maximization
in the last display into two intervals $\theta_{\mathsf{u}}\in[\delta,1]$
and $\theta_{\mathsf{u}}\in[0,\delta]$. The maximum over the first
interval, $\theta_{\mathsf{u}}\in[\delta,1]$ is bounded as: 
\begin{align}
 & M^{2}\cdot\max_{\theta_{\mathsf{u}}\in[\delta,1],\;\theta_{\mathsf{e}}\in[0,1-\theta_{\mathsf{u}}]}\P\left[|{\cal M}_{\mathsf{e}}|\geq\theta_{\mathsf{e}}M,\;|{\cal M}_{\mathsf{u}}|\geq\theta_{\mathsf{u}}M\right]\cdot e^{-M\cdot G_{R'}(\theta_{\mathsf{e}},\theta_{\mathsf{u}})+\log M}\nonumber \\
 & \leq M^{3}\cdot\P\left[|{\cal M}_{\mathsf{u}}|\geq\delta M\right]\cdot1\\
 & \leq M^{3}\cdot\exp\left[-c(1+o(1))\cdot(1-\sqrt{2\tau})\delta NL^{\zeta}\right],\label{eq: bounding for large number of undetected errors}
\end{align}
where we have utilized Lemma \ref{lem: erausre large deviations with high threshold}.
Similarly, the maximum over the second interval, $\theta_{\mathsf{u}}\in[0,\delta]$,
is bounded as: 
\begin{align}
 & M^{2}\cdot\max_{\theta_{\mathsf{u}}\in[0,\delta],\;\theta_{\mathsf{e}}\in[0,1-\theta_{\mathsf{u}}]}\P\left[|{\cal M}_{\mathsf{e}}|\geq\theta_{\mathsf{e}}M,\;|{\cal M}_{\mathsf{u}}|\geq\theta_{\mathsf{u}}M\right]\cdot e^{-M\cdot G_{R'}(\theta_{\mathsf{e}},\theta_{\mathsf{u}})+\log M}\nonumber \\
 & \leq M^{3}\cdot\max_{\theta_{\mathsf{e}}\in[0,1]}\P\left[|{\cal M}_{\mathsf{e}}|\geq\theta_{\mathsf{e}}M\right]\cdot e^{-M\cdot G_{R'}(\theta_{\mathsf{e}},\delta)},\label{eq: bounding for small number of undetected errors}
\end{align}
where the last inequality follows since $\theta_{\mathsf{u}}\to G_{R'}(\theta_{\mathsf{e}},\theta_{\mathsf{u}})$
is monotonic decreasing. Next, we will evaluate the maximum in \eqref{eq: bounding for small number of undetected errors}
over $\theta_{\mathsf{e}}\in[0,1-\delta]$ instead of $\theta_{\mathsf{e}}\in[0,1]$.
We will eventually take the limit $\delta\downarrow0$, and so the
continuity of our exponential bounds implies that the maximum over
interval $\theta_{\mathsf{e}}\in[1-\delta,1]$ can be ignored. To
further bound this term using Lemma \ref{lem: erausre large deviations with high threshold},
and to obtain the bound on the ensemble average error probability,
we separate the analysis to $N/M=\Theta(1)$ and $N/M=\omega(1)$.

\textbf{Case $N/M=\Theta(1)$: }In this case, Lemma \ref{lem: erausre large deviations with high threshold}
yields 
\begin{align}
-\log\overline{\pe}({\cal C}_{M},\mathsf{D}_{M}) & \geq M\cdot\lim_{\delta\downarrow0}\min\Bigg\{ c(1+o(1))\cdot(1-\sqrt{2\tau})\delta\frac{N}{M}L^{\zeta},\nonumber \\
 & \hphantom{===}\min_{\theta_{\mathsf{e}}\in[0,1-\delta]}d_{b}\left(\theta_{\mathsf{e}}||e^{-\varphi_{\tau}\frac{N}{M}}\right)\cdot\I\{\theta_{\mathsf{e}}\geq e^{-\varphi_{\tau}\frac{N}{M}}\}+G_{R'}(\theta_{\mathsf{e}},\delta)\Bigg\}-4\log M.
\end{align}
Consider the outer minimization between two terms for any $\delta>0$:
While the first term in the minimization is
\[
c(1+o(1))\cdot(1-\sqrt{2\tau})\delta\frac{N}{M}L^{\zeta}=\Theta(\log^{\zeta}M)=\omega(1),
\]
the second term is 
\[
\min_{\theta_{\mathsf{e}}\in[0,1-\delta]}d_{b}\left(\theta_{\mathsf{e}}||e^{-\varphi_{\tau}\frac{N}{M}}\right)\cdot\I\{\theta_{\mathsf{e}}\geq e^{-\varphi_{\tau}\frac{N}{M}}\}+G_{R'}(\theta_{\mathsf{e}},\delta)=O(1),
\]
as can be obtained by the choice $\theta_{\mathsf{e}}=1-\delta$ (which
yields $G_{R'}(1-\delta,\delta)=0$, and is possibly a sub-optimal
choice). Thus, for all large enough $M$, the minimum between the
two terms is obtained by the second term. The exponential bound on
the ensemble average error probability is obtained by solving (up
to negligible terms) 
\[
M\cdot\min_{\theta_{\mathsf{e}}\in[0,1-\delta]}d_{b}\left(\theta_{\mathsf{e}}||e^{-\varphi_{\tau}\frac{N}{M}}\right)\cdot\I\{\theta_{\mathsf{e}}\geq e^{-\varphi_{\tau}\frac{N}{M}}\}+\left[(R_{b}\beta-1)(1-\theta_{\mathsf{e}}-\delta)\log M-LR'\right]_{+}
\]
and then taking the limit $\delta\downarrow0$.\textbf{ }The last
minimization has the form 
\[
L\cdot\left[\min_{\theta_{\mathsf{e}}\in[0,1-\delta]}f(\theta_{\mathsf{e}})+[g(\theta_{\mathsf{e}})-R']_{+}\right],
\]
where 
\[
f(\theta_{\mathsf{e}})\dfn\frac{1}{L}d_{b}\left(\theta_{\mathsf{e}}||e^{-\varphi_{\tau}\frac{N}{M}}\right)\cdot\I\{\theta_{\mathsf{e}}\geq e^{-\varphi_{\tau}\frac{N}{M}}\}
\]
is a nonnegative, monotonically increasing, and convex function, and
\[
g(\theta_{\mathsf{e}})\dfn(R_{b}-1/\beta)(1-\theta_{\mathsf{e}}-\delta)
\]
is a nonnegative, monotonically decreasing and linear (and hence also
convex) function. This minimization is solved by a standard procedure
which involves finding the minimizer $\tilde{\theta}_{\mathsf{e}}$
for $R'=0$, the critical rate $R_{\text{cr}}$ which is the minimal
rate in which the clipping is active for $\tilde{\theta}_{\mathsf{e}}$,
i.e., $R_{\text{cr}}$ is defined as the unique rate satisfying $g(\tilde{\theta}_{\mathsf{e}})=R_{\text{cr}}$,
and for $R'>R_{\text{cr}}$ by setting $\tilde{\theta}_{\mathsf{e}}$
to be $\theta_{\mathsf{e},R'}$, which is defined as the unique parameter
satisfying $g(\theta)=R'$. The minimizer is then given by $\tilde{\theta}_{\mathsf{e}}$
for $0<R'<R_{\text{cr}}$, which yields the linear, unity-slope part
of the error exponent function, and by $\theta_{\mathsf{e},R'}$ for
$R'>R_{\text{cr}}$ which is curved part of the error exponent function.
The details appear in Proposition \ref{prop:minimization of error exponents},
Appendix \ref{sec:A-Minimization-Problem}.

According to the above procedure, we first find $\tilde{\theta}_{\mathsf{e}}$,
the minimizer of at zero rate $R=0$:
\[
\tilde{\theta}_{\mathsf{e}}\dfn\argmin_{\theta_{\mathsf{e}}\in[0,1-\delta]}d_{b}\left(\theta_{\mathsf{e}}||e^{-\varphi_{\tau}\frac{N}{M}}\right)\cdot\I\{\theta_{\mathsf{e}}\geq e^{-\varphi_{\tau}\frac{N}{M}}\}+(R_{b}\beta-1)(1-\theta_{\mathsf{e}}-\delta)\log M.
\]
The minimum value is $\Theta(1)$ as attained by the possibly sub-optimal
choice $\theta_{\mathsf{e}}=1-\delta$. Thus, it must hold that for
the minimizer $\tilde{\theta}_{\mathsf{e}}$ 
\[
(R_{b}\beta-1)(1-\tilde{\theta}_{\mathsf{e}}-\delta)\log M=O(1),
\]
and so $\tilde{\theta}_{\mathsf{e}}=1-\delta-c_{M}$ for some $c_{M}=O(\frac{1}{\log M})$.
Furthermore, since $\delta>0$ will be chosen to arbitrary small,
we may further restrict to $1-2\delta>\max_{\tau\in[0,1/2]}e^{-\varphi_{\tau}\frac{N}{M}}$.
For such restricted $\delta$, it holds that $\I\{1-\delta-c_{M}\geq e^{-\varphi_{\tau}\frac{N}{M}}\}=1$
for all large enough $M$, and the minimal value is then obtained
for $c_{M}$ chosen as 
\begin{align}
 & \min_{c_{M}\colon c_{M}=O(\frac{1}{\log M})}d_{b}\left(1-\delta-c_{M}||e^{-\varphi_{\tau}\frac{N}{M}}\right)+(R_{b}\beta-1)c_{M}\log M\nonumber \\
 & =d_{b}\left(1-\delta||e^{-\varphi_{\tau}\frac{N}{M}}\right)+O(c_{M}\log M).
\end{align}
The critical rate $R_{\text{cr}}$ is then 
\[
R_{\text{cr}}=g(\tilde{\theta}_{\mathsf{e}})=(R_{b}-1/\beta)c_{M}=o(1).
\]
Thus, for any $R'>0$, and all $M$ sufficiently large, the minimizing
$\theta$ is $\theta_{\mathsf{e},R'}$ (and so we are at the curved
part of the random coding exponent function), which is the solution
to 
\[
(R_{b}\beta-1)(1-\theta_{\mathsf{e},R'}-\delta)\log M=LR',
\]
or, as $L=\beta\log M$, 
\[
\theta_{\mathsf{e},R'}=1-\delta-\frac{R'}{R_{b}-1/\beta}.
\]
The resulting bound on the random coding error exponent is then
\begin{align}
-\log\overline{\pe}({\cal C}_{M},\mathsf{D}_{M}) & \geq M\cdot\lim_{\delta\downarrow0}d_{b}\left(\theta_{\mathsf{e},R'}||e^{-\varphi_{\tau}\frac{N}{M}}\right)\cdot\I\{\theta_{\mathsf{e},R'}\geq e^{-\varphi_{\tau}\frac{N}{M}}\}-O\left(\log M\right)\\
 & =-O\left(\log M\right)+\begin{cases}
M\cdot d_{b}\left(\theta_{\mathsf{e},R'}||e^{-\varphi_{\tau}\frac{N}{M}}\right), & 1-\frac{R'}{R_{b}-1/\beta}\geq e^{-\varphi_{\tau}\frac{N}{M}}\\
0, & 1-\frac{R'}{R_{b}-1/\beta}\leq e^{-\varphi_{\tau}\frac{N}{M}}
\end{cases}.
\end{align}
The condition for a positive exponent in the last display is 
\[
R'\leq(R_{b}-1/\beta)\left(1-e^{-\varphi_{\tau}\frac{N}{M}}\right).
\]
This exponent is maximized by minimizing $e^{-\varphi_{\tau}\frac{N}{M}}$,
that is maximizing $\varphi_{\tau}=-\frac{1}{N/M}\log\P[\tilde{S}_{m}\leq T_{\tau}]$.
This maximum is clearly obtained by minimizing $T_{\tau}=\frac{N}{M}(1-\sqrt{2\tau})$,
or, equivalently, taking the limit $\tau\uparrow1/2$. Since $N/M=\Theta(1)$
there exists $\delta^{*}>0$ such that for all $\delta\in(0,\delta^{*})$
it holds that $T_{1/2-\delta}<1$. So, for such $\delta$ 
\begin{align}
\varphi_{1/2-\delta} & =-\frac{1}{N/M}\log\P[\tilde{S}_{m}\leq T_{1/2-\delta}]\\
 & =-\frac{1}{N/M}\log\P\left[\tilde{S}_{m}=0\right]=1.
\end{align}
Hence, by taking $\delta\downarrow0$, which specifically results
$\delta<\delta^{*}$, yields the bound 
\begin{align}
-\liminf_{M\to\infty}\frac{1}{M}\log\overline{\pe}({\cal C}_{M},\mathsf{D}_{M}) & \geq\liminf_{M\to\infty}\begin{cases}
d_{b}\left(1-\frac{R'}{R_{b}-1/\beta}||e^{-\frac{N}{M}}\right), & R'\leq(R_{b}-1/\beta)\left(1-e^{-\frac{N}{M}}\right)\\
0, & R'>(R_{b}-1/\beta)\left(1-e^{-\frac{N}{M}}\right)
\end{cases}\\
 & \trre[=,*]\begin{cases}
d_{b}\left(1-\frac{R}{R_{b}-1/\beta}||e^{-\frac{N}{M}}\right), & R\leq(R_{b}-1/\beta)\left(1-e^{-\frac{N}{M}}\right)\\
0, & R>(R_{b}-1/\beta)\left(1-e^{-\frac{N}{M}}\right)
\end{cases},
\end{align}
where $(*)$ follows from $R'=R+\frac{1}{L}$, and a first-order Taylor
approximation of $d_{b}(\theta||e^{-\frac{N}{M}})$ around $\theta=1-\frac{R}{R_{b}-1/\beta}$.
The claimed bound then follows.

\textbf{Case $N/M=\omega(1)$: }As in the previous case, we utilize
Lemma \ref{lem: erausre large deviations with high threshold}, and
continue bounding \eqref{eq: random coding bound with sequencing proof}
separately for the two intervals $\theta_{\mathsf{u}}\in[0,\delta]$
and $\theta_{\mathsf{u}}\in[\delta,1]$. The bound \eqref{eq: bounding for large number of undetected errors}
remains the same, and this leads to the bound 
\begin{align}
 & -\log\overline{\pe}({\cal C}_{M},\mathsf{D}_{M})\nonumber \\
 & \geq ML\cdot\lim_{\delta\downarrow0}\min\Bigg\{ c\cdot(1+o(1))(1-\sqrt{2\tau})\delta\frac{N}{ML^{1-\zeta}},\nonumber \\
 & \hphantom{===}\min_{\theta_{\mathsf{e}}\in[0,1-\delta]}\theta_{\mathsf{e}}\tau\frac{N}{ML}\cdot[1+o(1)]\I\{\theta_{\mathsf{e}}\geq e^{-\tau\frac{N}{M}}\}+\frac{G_{R'}(\theta_{\mathsf{e}},\delta)}{L}\Bigg\}+O\left(\log M\right).
\end{align}
As in the previous case, the first term in the outer minimization
(which is $\Theta(\frac{N}{ML^{1-\zeta}})$) is larger than the second
term (which is $O(\frac{N}{ML})$). Hence, the minimum is obtained
by the second term for all large enough $M$, and so the required
exponential bound is obtained by solving 
\[
\min_{\theta_{\mathsf{e}}\in[0,1-\delta]}\theta_{\mathsf{e}}\tau\frac{N}{ML}\cdot[1+o(1)]\I\{\theta_{\mathsf{e}}\geq e^{-\tau\frac{N}{M}}\}+\left[(R_{b}-1/\beta)(1-\theta_{\mathsf{e}}-\delta)-R'\right]_{+}.
\]
We follow the same procedure as in the $N/M=\Theta(1)$ case. For
$R'=0$, it is required to solve 
\[
\min_{\theta_{\mathsf{e}}\in[0,1-\delta]}\theta_{\mathsf{e}}\tau\frac{N}{ML}\I\{\theta_{\mathsf{e}}\geq e^{-\tau\frac{N}{M}}\}+(R_{b}-1/\beta)(1-\theta_{\mathsf{e}}-\delta).
\]
We next consider three sub-cases:
\begin{itemize}
\item Suppose that $\frac{N}{ML}=o(1)$. Then the term $(R_{b}-1/\beta)(1-\theta_{\mathsf{e}}-\delta)$
asymptotically dominates and $\tilde{\theta}_{\mathsf{e}}\to1-\delta$.
The critical rate is $R_{\text{cr}}=o(1)$, and for any $R'>0$ and
all sufficiently large $M$ so that $R'>R_{\text{cr}}$
\[
\theta_{\mathsf{e},R'}=1-\delta-\frac{R'}{R_{b}-1/\beta}.
\]
In addition, for $R'\leq(1-\delta)(R_{b}-1/\beta)$ and all large
enough $M$, it holds that $\theta_{\mathsf{e},R'}>e^{-\tau\frac{N}{M}}=o(1)$
and so the exponent bound is 
\[
\left[1-\delta-\frac{R'}{R_{b}-1/\beta}\right]\tau\frac{N}{ML}.
\]
Maximizing the bound by taking $\tau=1/2-\delta$ using $R'=R+\frac{1}{L}$,
for all $R\leq R_{b}-1/\beta$, and then taking $\delta\downarrow0$
yields the bound
\[
-\liminf_{N\to\infty}\frac{1}{N}\log\overline{\pe}({\cal C}_{M},\mathsf{D}_{M})\geq\frac{1}{2}\left[1-\frac{R}{R_{b}-1/\beta}\right].
\]
\item Suppose that $\frac{N}{ML}=a$ where $a=\Theta(1)$. At $R'=0$ the
minimization problem is 
\begin{align}
 & \min_{\theta_{\mathsf{e}}\in[0,1-\delta]}\theta_{\mathsf{e}}a\tau\I\{\theta_{\mathsf{e}}\geq e^{-\tau\frac{N}{M}}\}+(R_{b}-1/\beta)(1-\theta_{\mathsf{e}}-\delta)\nonumber \\
 & =\min_{\theta_{\mathsf{e}}\in[0,1-\delta]}\theta_{\mathsf{e}}\left[a\tau\I\{\theta_{\mathsf{e}}\geq e^{-\tau\frac{N}{M}}\}-(R_{b}-1/\beta)\right]+(R_{b}-1/\beta)(1-\delta).
\end{align}
 The minimum is attained at 
\[
\tilde{\theta}_{\mathsf{e}}=\begin{cases}
0, & a>\frac{R_{b}-1/\beta}{\tau}\\
1-\delta, & a<\frac{R_{b}-1/\beta}{\tau}
\end{cases},
\]
the values are, respectively, $(R_{b}-1/\beta)(1-\delta)$ and $(1-\delta)a\tau$,
and the critical rates are, respectively, $R_{\text{cr}}=(R_{b}-1/\beta)(1-\delta)$
and $R_{\text{cr}}=0$. In the former case, $\tilde{\theta}_{\mathsf{e}}=0$,
and so the exponent is a linear unity-slope function given by $(R_{b}-1/\beta)(1-\delta)-R'$.
In the latter case, $\tilde{\theta}_{\mathsf{e}}=1-\delta$, the exponent
for all $R'>0$ is found by setting 
\[
\theta_{\mathsf{e},R'}=1-\delta-\frac{R'}{R_{b}-1/\beta}
\]
for which $\I\{\theta_{\mathsf{e},R'}\geq e^{-\tau\frac{N}{M}}\}=1$
for all $M$ sufficiently large. This yields the exponent 
\[
\theta_{\mathsf{e},R'}\tau a\cdot[1+o(1)]\cdot\I\{\theta_{\mathsf{e},R'}\geq e^{-\tau\frac{N}{M}}\}=[1+o(1)]\cdot\tau a\left[1-\delta-\frac{R'}{R_{b}-1/\beta}\right].
\]
Maximizing the bounds in both cases by setting $\tau=1/2-\delta$,
and then taking $\delta\downarrow0$ we obtain the exponent bound
\[
-\liminf_{N\to\infty}\frac{1}{N}\log\overline{\pe}({\cal C}_{M},\mathsf{D}_{M})\geq ML\left[\frac{a}{2}\wedge\left(R_{b}-1/\beta\right)\right]\left[1-\frac{R}{R_{b}-1/\beta}\right].
\]
\item Suppose that $\frac{N}{ML}=\omega(1)$. Then, at $R'=0$ the minimization
problem is 
\[
\min_{\theta_{\mathsf{e}}\in[0,1]}\theta_{\mathsf{e}}\frac{N}{ML}\tau\cdot\I\{\theta_{\mathsf{e}}\geq e^{-\tau\frac{N}{M}}\}+(R_{b}-1/\beta)(1-\theta_{\mathsf{e}}-\delta).
\]
The minimizer is thus $\tilde{\theta}_{\mathsf{e}}=e^{-\tau\frac{N}{M}}+o(1)=o(1)$
and at $R'=0$ the exponent bound is 
\[
(R_{b}-1/\beta)(1-e^{-\tau\frac{N}{M}}-\delta)=(R_{b}-1/\beta)(1-\delta)+o(1).
\]
The critical rate is 
\[
R_{\text{cr}}=(R_{b}-1/\beta)(1-e^{-\tau\frac{N}{M}}-\delta)=(R_{b}-1/\beta)(1-\delta)+o(1).
\]
 Setting again $\tau=1/2-\delta$, and then taking $\delta\downarrow0$
and using $R'=R+\frac{1}{L}$ leads to the exponential bound 
\[
-\liminf_{N\to\infty}\frac{1}{N}\log\overline{\pe}({\cal C}_{M},\mathsf{D}_{M})\geq ML\left[R_{b}-1/\beta-R\right].
\]
\end{itemize}
Summarizing all possible cases leads to the claimed exponential bounds.
\end{IEEEproof}

\subsection{Expurgated Bound Analysis \label{subsec:Expurgated-Bound-Analysis}}

In this section, we improve the random coding bound in some regime
of $N$ via an expurgation argument. Specifically, we establish the
following:
\begin{prop}[Expurgated bound]
\label{prop: index molecule expurgated bound}Let an inner code be
given by $\{{\cal B}_{m}^{(L)}\}\subset{\cal X}^{L}$ for $m\in[M]$,
and let $\mathsf{D}_{b}$ be a decoder which satisfy the assumptions
on the inner code at rate $R_{b}$. Then, there exists a sequence
of coded-index based codebooks $\{{\cal C}_{M}\}$ and corresponding
decoders (as described in Sec. \ref{subsec:Formulation-of-the}) such
that if $N/(ML)>4(R_{b}-1/\beta)$ then 
\[
\liminf_{N\to\infty}-\log\pe({\cal C}_{M},\mathsf{D}_{M})\geq\frac{N}{4}\left[1-\frac{R}{R_{b}-1/\beta}\right],
\]
for any $R<R_{b}-1/\beta$.
\end{prop}
We follow the expurgation argument of \cite[Appendix]{merhav2014list}.
To this end, we define the \emph{distance enumerator} of the $j$th
codeword, $j\in|{\cal C}|$, of a coded-index based codebook ${\cal C}$
at relative distance $\gamma\in\frac{1}{M}[M+1]$, as 
\[
\mathscr{N}_{j}({\cal C},\gamma)\dfn\left|\left\{ \tilde{j}\neq j\colon\rho(x^{LM}(j),x^{LM}(\tilde{j}))=\gamma M\right\} \right|.
\]
We begin by proving the existence of a codebook with a specified distance
spectrum. 
\begin{lem}
\label{lem: distance spectrum expurgation}Let rate $R$ be given,
and assume that $\eta\equiv\eta_{M}=4/(M\beta)$. Then, there exists
a codebook ${\cal C}^{*}$ of rate $R-\Theta(\frac{1}{M})$ so that
\[
\mathscr{N}_{j}({\cal C}^{*},\gamma)\begin{cases}
\leq\exp\left[ML\cdot\left(R-(1-\gamma)(R_{b}-1/\beta)\right)\right], & \gamma\geq1-\frac{R-\eta}{R_{b}-1/\beta}\\
=0, & \gamma<1-\frac{R-\eta}{R_{b}-1/\beta}
\end{cases}.
\]
\end{lem}
\begin{IEEEproof}
The proof is by random selection over the coded-index based random
coding ensemble (Definition \ref{def:RCE}). We begin by evaluating
the expected value of $\mathscr{N}_{j}({\cal C},\gamma)$ over the
ensemble. By symmetry and the uniformity of the random selection of
the codebook, it is clear that we may assume w.l.o.g. that $j=0$,
and that $x^{LM}(0)$ is arbitrary. By the definition of the random
coding ensemble, the $m$th molecule of each codeword $x^{LM}(j)$
is chosen uniformly at random from ${\cal B}_{m}^{(L)}$ whose cardinality
is $|{\cal B}_{m}^{(L)}|=\frac{e^{R_{b}L}}{M}=M^{R_{b}\beta-1}$.
Thus, the probability that $x^{LM}(0)$ and $x^{LM}(1)$ (or any other
$j>0$) are equal in their $m$th molecule is $M^{1-R_{b}\beta}=o(1)$
(recall the assumption $R_{b}\beta>1$). Thus, the expected number
of identical molecules in $x^{LM}(0)$ and $x^{LM}(1)$ is $M^{2-R_{b}\beta}=o(M)$.
So, the probability that there are $\gamma M$ identical molecules
in $x^{LM}(0)$ and $x^{LM}(1)$ is a large-deviations event. Chernoff's
bound then implies 
\begin{align}
\P\left[\rho(x^{LM}(0),X^{LM}(1))=\gamma M\right] & =\P\left[\text{Binomial}(M,M^{1-R_{b}\beta})=(1-\gamma)M\right]\\
 & \leq\exp\left[-M\cdot d_{b}\left(1-\gamma\mid\mid M^{1-R_{b}\beta}\right)\right]\\
 & \leq\exp\left[-ML\cdot[1+o(1)]\cdot(1-\gamma)(R_{b}-1/\beta)\right],
\end{align}
where the last inequality is by the asymptotic expansion of the binary
KL divergence $d_{b}(a||b)=-[1+o(1)]\cdot a\log b$ (see Proposition
\ref{prop: Asymptotic expansions of the binary KL divergence} in
Appendix \ref{sec:Asymptotic-Expansions-of}). Then, by linearity
of expectation 
\begin{align}
\E\left[\mathscr{N}_{0}({\cal C},\gamma)\right] & =(|{\cal C}|-1)\cdot\P\left[\rho(x^{LM}(1),X^{LM}(2))=\gamma M\right]\\
 & \leq\exp\left[ML\cdot\left(R-[1+o(1)](1-\gamma)(R_{b}-1/\beta)\right)\right].
\end{align}
Similar bound holds for any $j>0$. Then, by the union bound and Markov's
inequality,
\begin{align}
 & \P\left[\bigcup_{\gamma\in\frac{1}{M}[M+1]}\left\{ \frac{1}{|{\cal C}|}\sum_{j\in[|{\cal C}|]}\mathscr{N}_{j}({\cal C},\gamma)\geq\exp\left[ML\cdot\left(R-(1-\gamma)(R_{b}-1/\beta)+\eta/2\right)\right]\right\} \right]\nonumber \\
 & \leq\sum_{\gamma\in\frac{1}{M}[M+1]}\P\left[\left\{ \frac{1}{|{\cal C}|}\sum_{j\in[|{\cal C}|]}\mathscr{N}_{j}({\cal C},\gamma)\geq\exp\left[ML\cdot\left(R-(1-\gamma)(R_{b}-1/\beta)+\eta/2\right)\right]\right\} \right]\\
 & \leq M\cdot e^{-ML\eta/2}=e^{-ML\left(\eta/2-\frac{\log M}{ML}\right)}=e^{-ML\eta/4}=\frac{1}{M},
\end{align}
where the last inequality follows from the choice $\eta=4/(M\beta)$.
So, with probability larger than $1-\frac{1}{M}$
\begin{equation}
\frac{1}{|{\cal C}|}\sum_{j\in[|{\cal C}|]}\mathscr{N}_{j}({\cal C},\gamma)\leq\exp\left[ML\cdot\left(R-(1-\gamma)(R_{b}-1/\beta)+\eta/2\right)\right]\label{eq: distance specturm random coding property}
\end{equation}
holds for all $\gamma\in\frac{1}{M}[M+1]$. Let ${\cal C}'$ be a
code which satisfies the property \eqref{eq: distance specturm random coding property}
for all $\gamma\in\frac{1}{M}[M+1]$ whose cardinality is $e^{MLR}$.
Then, for any given $\gamma\in\frac{1}{M}[M+1]$ there must exist
a sub-codebook ${\cal C}''(\gamma)$ of cardinality larger than $(1-e^{-ML\eta/2})\cdot e^{MLR}$
such that 
\begin{equation}
\mathscr{N}_{j}({\cal C}''(\gamma),\gamma)\leq\exp\left[ML\cdot\left(R-(1-\gamma)(R_{b}-1/\beta)+\eta\right)\right]\label{eq: distance spectrum for an expurgated codebook given gamma}
\end{equation}
holds for any $j\in[|{\cal C}''(\gamma)|]$. Indeed, assume by contradiction
that this is not the case. Then, there exists more than $e^{-ML\eta/2}\cdot e^{MLR}$
codewords for which $\mathscr{N}_{j}({\cal C}',\gamma)$ satisfies
the opposite inequality to \eqref{eq: distance spectrum for an expurgated codebook given gamma}.
Summing $\frac{1}{|{\cal C}|}\mathscr{N}_{j}({\cal C}',\gamma)$ over
these codewords will then lead to a value larger than the right-hand
side (r.h.s.) of \eqref{eq: distance specturm random coding property},
which is a contradiction. Thus, ${\cal C}'''\dfn\cap_{\gamma\in\frac{1}{M}[M+1]}{\cal C}''(\gamma)$
satisfies that 
\begin{equation}
\mathscr{N}_{j}({\cal C}''',\gamma)\leq\exp\left[ML\cdot\left(R-(1-\gamma)(R_{b}-1/\beta)+\eta\right)\right]\label{eq: final bound on enumerator in the proof}
\end{equation}
for all $j\in[|{\cal C}'''|]$ and $\gamma\in\frac{1}{M}[M+1]$. In
addition, 
\begin{align}
|{\cal C}\backslash{\cal C}'''| & =\left|{\cal C}\backslash\cap_{\gamma\in\frac{1}{M}[M+1]}{\cal C}''(\gamma)\right|\\
 & =\left|\cup_{\gamma\in\frac{1}{M}[M+1]}{\cal C}\backslash{\cal C}''(\gamma)\right|\\
 & \trre[\leq,a]\sum_{\gamma\in\frac{1}{M}[M+1]}\left|{\cal C}\backslash{\cal C}''(\gamma)\right|\\
 & \trre[\leq,b]\sum_{\gamma\in\frac{1}{M}[M+1]}e^{-ML\eta/2}e^{MLR}\\
 & \leq M\cdot e^{-ML\eta/2}e^{MLR},
\end{align}
where $(a)$ follows from the union bound, $(b)$ follows since $|{\cal C}''(\gamma)|\geq(1-e^{-ML\eta/2})\cdot e^{MLR}$
for all $\gamma\in\frac{1}{M}[M+1]$. Hence, 
\[
|{\cal C}'''|\geq(1-Me^{-ML\eta/2})\cdot e^{MLR}=(1-\frac{1}{M})\cdot e^{MLR}\geq e^{ML(R-\Theta(1/(ML)))},
\]
for all large enough $M$. The proof is completed by noting that since
the left-hand side of \eqref{eq: final bound on enumerator in the proof}
is integer, it must be zero whenever the r.h.s. is strictly less than
$1$.
\end{IEEEproof}
We next prove Proposition \ref{prop: index molecule expurgated bound}:
\begin{IEEEproof}[Proof of Proposition \ref{prop: index molecule expurgated bound}]
Let $\eta=\frac{4}{\beta M}$ be given and assume that $M$ is large
enough so that $\eta\in(0,R)$. Further let ${\cal C}^{*}$ be a code
as guaranteed by Lemma \ref{lem: distance spectrum expurgation} of
rate $R'=R-\eta$, and let $\mathsf{D}$ be the corresponding decoder,
as described in Sec. \ref{subsec:Formulation-of-the}. By the union
bound, for any $j\in[|{\cal C}^{*}|]$
\begin{align}
\pe({\cal C}^{*},\mathsf{D}\mid x^{LM}(j)) & \leq\sum_{\tilde{j}\in[|{\cal C}^{*}|]\backslash\{j\}}\P\left[\rho(\hat{x}^{LM},x^{LM}(\tilde{j}))<\rho(\hat{x}^{LM},x^{LM}(j))\right]\\
 & \leq\sum_{\gamma\in\frac{1}{M}[M+1]}\sum_{\tilde{j}\in[|{\cal C}^{*}|]\backslash\{j\}\colon\rho(x^{LM}(j),x^{LM}(\tilde{j}))=\gamma M}\P\left[\rho(\hat{x}^{LM},x^{LM}(\tilde{j}))<\rho(\hat{x}^{LM},x^{LM}(j))\right].
\end{align}
We next bound the probability in the above summation for some arbitrary
pair of codewords $j\neq\tilde{j}$ for which $\rho(x^{LM}(j),x^{LM}(\tilde{j}))=\gamma M$.
For the sake of this bound, we assume w.l.o.g. that this pair of codewords
has different molecules in the set $[\gamma M]$. In Sec. \ref{subsec:Error-Events-for}
we have defined the cardinality of erasure and undetected error events
for the entire set of molecules $[M]$ {[}recall the definitions of
${\cal M}_{\mathsf{e}}$ and ${\cal M}_{\mathsf{u}}$ in \eqref{eq: erasure set}
and \eqref{eq: undetected error set}{]}, and here, we consider, in
lieu of ${\cal M}_{\mathsf{e}}$ and ${\cal M}_{\mathsf{u}}$, a similar
sets of indices which are restricted to $[\gamma M]$, to wit, 
\begin{align}
{\cal M}_{\mathsf{e}}^{(\gamma)} & \dfn\left\{ m\in[\gamma M]\colon\hat{x}_{m}^{L}=\mathsf{e}\right\} ,\label{eq: erasure set modified}\\
{\cal M}_{\mathsf{u}}^{(\gamma)} & \dfn\left\{ m\in[\gamma M]\colon\hat{x}_{m}^{L}\neq\mathsf{e},\;\hat{x}_{m}^{L}\neq x_{m}^{L}(j)\right\} .\label{eq: undetected error set modified}
\end{align}
Analogously to Lemma \ref{lem:Bounds on the size of erasure and undetected sets},
it then holds that
\[
|{\cal M}_{\mathsf{e}}^{(\gamma)}|+|{\cal M}_{\mathsf{u}}^{(\gamma)}|\leq|{\cal M}_{\text{sam}}^{(\gamma)}|+\left(1+\frac{2}{1-\sqrt{2\tau}}\cdot\frac{M}{N}\right)K,
\]
where 
\[
{\cal M}_{\text{sam}}^{(\gamma)}\dfn\left\{ m\in[\gamma M]\colon S_{m}<T_{\tau}\right\} ,
\]
and $K$ is defined, exactly as in the random coding analysis of Sec.
\ref{subsec:Random-Coding-Analysis}, as the total number of molecules
which were erroneously sequenced (even those in $[M]\backslash[\gamma M]$).
With these definitions, we may further upper bound 
\begin{align}
 & \pe({\cal C}^{*},\mathsf{D}\mid x^{LM}(j))\nonumber \\
 & \leq\sum_{\gamma\in\frac{1}{M}[M+1]}\mathscr{N}_{j}({\cal C}^{*},\gamma)\cdot\P\left[|{\cal M}_{\mathsf{e}}^{(\gamma)}|+|{\cal M}_{\mathsf{u}}^{(\gamma)}|\geq\frac{1}{2}\gamma M\right]\\
 & \leq\sum_{\gamma\in\frac{1}{M}[M+1]}\mathscr{N}_{j}({\cal C}^{*},\gamma)\cdot\P\left[|{\cal M}_{\text{sam}}^{(\gamma)}|+\left(1+\frac{2}{1-\sqrt{2\tau}}\cdot\frac{M}{N}\right)K\geq\frac{1}{2}\gamma M\right]\\
 & \leq\sum_{\gamma\in\frac{1}{M}[M+1]}\mathscr{N}_{j}({\cal C}^{*},\gamma)\cdot\sum_{\sigma\in\frac{1}{M}\cdot[M+1],\;\kappa\in\frac{1}{N}\cdot[N+1]\colon\;\sigma+\kappa\left(\frac{N}{M}+\frac{2}{1-\sqrt{2\tau}}\right)\geq\frac{\gamma}{2}}\P\left[|{\cal M}_{\text{sam}}^{(\gamma)}|\geq\sigma M,\;K\geq\kappa N\right]\\
 & \trre[\leq,*]\sum_{\gamma\in\frac{1}{M}[M+1]}\mathscr{N}_{j}({\cal C}^{*},\gamma)\cdot\sum_{\sigma\in\frac{1}{M}\cdot[M+1],\;\kappa\in\frac{1}{N}\cdot[N+1]\colon\;\sigma+\kappa\left(\frac{N}{M}+\frac{2}{1-\sqrt{2\tau}}\right)\geq\frac{\gamma}{2}}\P\left[|{\cal M}_{\text{sam}}^{(\gamma)}|\geq\sigma M\right]\cdot\P\left[\tilde{K}\geq\kappa N\right],
\end{align}
where $(*)$ follows similarly to Lemma \ref{lem: total seqeuncing errors large deviations}.
We next bound $\mathscr{N}_{j}({\cal C}^{*},\gamma)$ by exploiting
the guarantees on ${\cal C}^{*}$ (from Lemma \ref{lem: distance spectrum expurgation}),
and bounds on the probabilities $\P[|{\cal M}_{\text{sam}}^{(\gamma)}|\geq\sigma M]$
and $\P[\tilde{K}\geq\kappa N]$. For the former, it can be easily
deduced that the bound of Lemma \ref{lem: amplification large deviations}
holds verbatim, and given by
\[
\P\left[|{\cal M}_{\text{sam}}^{(\gamma)}|\geq\sigma M\right]\leq4e^{-\sigma\tau N\cdot[1+o(1)]}
\]
for $\sigma\in(e^{-\tau\frac{N}{M}},1]$. This is because the reduction
in the randomness due to the change from a sum of $M$ r.v.'s to a
sum of $\gamma M$ r.v.'s is compensated by the increase in the relative
probability required to cross the threshold $\sigma M$. For the latter,
we again use Lemma \ref{lem: total seqeuncing errors large deviations}
verbatim for $\tilde{K}$ instead of $K$. 

We next plug in those bounds only the\textbf{ $N/M=\omega(1)$ }case,
as similar analysis for the other cases shows that there is no improvement
for the \textbf{$N/M=\Theta(1)$ }case. So, assuming \textbf{$N/M=\omega(1)$},
the bound is 
\begin{align}
 & -\log\pe({\cal C}^{*},\mathsf{D}\mid x^{LM}(j))\nonumber \\
 & \geq ML\cdot\min_{\gamma\in[1-\frac{R'}{R_{b}-1/\beta},1]}\min_{\sigma\in[0,1],\;\kappa\in[0,1]\colon\sigma+\kappa\left(\frac{N}{M}+\frac{2}{1-\sqrt{2\tau}}\right)\geq\frac{\gamma}{2}}\left\{ \sigma\tau\frac{N(1+o(1))}{ML}+c\cdot\kappa\frac{N}{ML}L^{\zeta}+(1-\gamma)(R_{b}-1/\beta)-R'\right\} \nonumber \\
 & \hphantom{==}-O\left(\log M\right)
\end{align}
for $\sigma\in(e^{-\tau\frac{N}{M}},1]$. Considering the inner minimization
for some given $\gamma$, the minimum must be attained for $\kappa=0$
since $L^{\zeta}=\omega(1)$. This leads to the bound 
\begin{align}
 & -\log\pe({\cal C}^{*},\mathsf{D}\mid x^{LM}(j))\nonumber \\
 & \geq ML\cdot\min_{\gamma\in[1-\frac{R'}{R_{b}-1/\beta},1]}\min_{\sigma\in[0,1]\colon\sigma\geq\frac{\gamma}{2}}\left\{ \sigma\tau\frac{N}{ML}N(1+o(1))+(1-\gamma)(R_{b}-1/\beta)-R'\right\} -O(\log M)\\
 & =\min_{\gamma\in[1-\frac{R'}{R_{b}-1/\beta},1]}\left\{ \frac{\gamma\tau}{2}\frac{N(1+o(1))}{ML}+(1-\gamma)(R_{b}-1/\beta)-R'\right\} -O(\log M)
\end{align}
for $\sigma\in(e^{-\tau\frac{N}{M}},1]$. It is evident from that
last bound that it is optimal to set $\tau\uparrow1/2$ and so we
continue analyzing the bound with this choice, by further restricting
to the case $\frac{N}{ML}>4(R_{b}-1/\beta)$ assumed in the statement
of the proposition. Then, the minimizer is obtained for $\gamma=1-\frac{R'}{R_{b}-1/\beta}$
and the non-vanishing term is 
\[
ML\cdot\min_{\gamma\in[1-\frac{R-\eta}{R_{b}-1/\beta},1]}\left\{ \gamma\cdot\frac{N(1+o(1))}{4ML}+(1-\gamma)(R_{b}-1/\beta)-R'\right\} \geq\left(1-\frac{R}{R_{b}-1/\beta}\right)\frac{N(1+o(1))}{4}-O\left(\frac{N}{M}\right)
\]
where we have used $R'=R-\eta$ and the assumption that $\eta=\Theta(1/M)$. 
\end{IEEEproof}

\section{Summary\label{sec:Summary}}

We have considered a simple and general coding scheme for the DNA
storage channel and analyzed its error probability. In the analysis
of this scheme and in our previous research \cite{weinberger2021DNA}
it was identified that sampling events dominate the error probability,
and so lowering the rate of sequencing errors in this scheme is of
secondary importance compared to proper sampling of all molecules.
This phenomenon resembles wireless fading communication channels \cite{tse2005fundamentals},
in which the transmitted signal experiences both fading (a random
gain, or multiplicative noise) and additive noise (typically assumed
to be Gaussian). Under common fading models such as Rayleigh, the
probability that the random gain is close to zero decays \emph{polynomially}
with the signal-to-noise ratio, and in this event the output signal
is ``lost'', and so an error is inevitable. This event dominates
the error probability, and as a result the error probability decays
much slower compared to fixed gain additive noise channels, in which
the decay rate is exponential (e.g., \cite[Ch. 7]{gallager1968information}).
Analogously, random sampling events, in which too many molecules are
under-sampled, dominates the error probability over the sequencing
errors.\footnote{At least in the analysis of this paper and that of \cite{weinberger2021DNA}.}.
Hence, since the error probability bound of this paper scales as $e^{-\Theta(N)}$,
and analogously to \emph{diversity }techniques \cite[Ch. 3]{tse2005fundamentals}
in wireless communication, the importance of increasing the coverage
depth $N$ as much as possible. 

Nonetheless, as discussed in \cite{shomorony2021dna}, future systems
will aim for faster and cheaper sequencing machines, which inevitably
will increase sequencing errors, which even include non-negligible
rate of deletions and insertions. Constructing \emph{practical} coding
methods and decoders for this channel, as was studied in \cite{lenz2020achievable}
and in this paper, is an important avenue for future research.

\section*{Acknowledgment}

Discussions with N. Merhav on this topic are acknowledged with gratitude.
Various comments and suggestions made by the anonymous referees have
significantly improved the manuscript, and are also acknowledged with
gratitude. 

\appendices{\numberwithin{equation}{section}}

\section{Asymptotic Expansion of the Binary KL Divergence\label{sec:Asymptotic-Expansions-of}}
\begin{prop}
\label{prop: Asymptotic expansions of the binary KL divergence}If
$a\in[0,1]$ and $b_{n}=o(1)$ then $d_{b}(a||b_{n})=-[1+o(1)]\cdot a\log b_{n}$.
\end{prop}
\begin{IEEEproof}
It holds that 
\[
a\log\frac{a}{b_{n}}=-[1+o(1)]\cdot a\log b_{n}=\omega(1)
\]
and using the expansion $\log(1+x)=x+\Theta(x^{2})$ we obtain that
\begin{align}
(1-a)\log\frac{(1-a)}{(1-b_{n})} & =(1-a)\log(1-a)-(1-a)\log(1-b_{n})\\
 & =(1-a)\log(1-a)+b_{n}(1-a)+\Theta(b_{n}^{2})\\
 & =[1+o(1)]\cdot(1-a)\log(1-a)\\
 & =\Theta(1).
\end{align}
The result then follows by adding both terms. 
\end{IEEEproof}

\section{A Minimization Problem in Random Coding Exponent Bounds \label{sec:A-Minimization-Problem}}

The following provides a general solution for a standard minimization
problem that occurs in the derivations of random coding bounds of
the error exponent (as developed by Csisz\'{a}r and K\"{o}rner \cite[Ch. 10]{csiszar2011information}). 
\begin{prop}
\label{prop:minimization of error exponents}Suppose that $f(\theta)$
and $g(\theta)$ are nonnegative and convex functions on $[0,1]$,
and that $f(\theta)$ is strictly increasing while $g(\theta)$ is
strictly decreasing. Let 
\[
\theta_{0}\in\argmin_{\theta\in[0,1]}f(\theta)+g(\theta),
\]
let $R_{\text{\emph{cr}}}\dfn g(\theta_{0})$, and let $\theta_{R}$
be defined by $g(\theta_{R})=R$ for $R>R_{\text{\emph{cr}}}$. Then,
\[
\min_{\theta\in[0,1]}f(\theta)+[g(\theta)-R]_{+}=\begin{cases}
f(\theta_{0})+g(\theta_{0})-R, & R\leq R_{\text{\emph{cr}}}\\
f(\theta_{R}), & R\geq R_{\text{\emph{cr}}}
\end{cases}.
\]
\end{prop}
\begin{IEEEproof}
For $R=0$ the minimum is clearly attained for $\theta_{0}$. Then,
for $R<R_{\text{cr}}=g(\theta_{0})$
\begin{align}
f(\theta_{0})+[g(\theta_{0})-R]_{+} & =f(\theta_{0})+g(\theta_{0})-R\\
 & =\min_{\theta\in[0,1]}f(\theta)+g(\theta)-R\\
 & \leq\min_{\theta\in[0,1]}f(\theta)+[g(\theta)-R]_{+}
\end{align}
and so $\theta_{0}$ is the minimizer for all $R\leq R_{\text{cr}}$.
For $R>R_{\text{cr}}$, let us write 
\begin{align}
 & \min_{\theta\in[0,1]}f(\theta)+[g(\theta)-R]_{+}\nonumber \\
 & =\min\Bigg\{\min_{\theta\in[0,1]\colon g(\theta)\leq R}f(\theta),\min_{\theta\in[0,1]\colon g(\theta)\geq R}f(\theta)+g(\theta)-R\Bigg\}.
\end{align}
Regarding the second inner minimization, that is constrained to $\{\theta\in[0,1]\colon g(\theta)\geq R\}$,
we may alternatively inspect the unconstrained minimization, and note
that $\theta_{0}$, the unconstrained minimizer, satisfies $g(\theta_{0})<R$.
The convexity of $f(\theta)+g(\theta)$ then implies that the solution
to the same minimization problem constrained to $\{\theta\in[0,1]\colon g(\theta)\geq R\}$,
that is
\[
\min_{\theta\in[0,1]\colon g(\theta)\geq R}f(\theta)+g(\theta)-R
\]
can be attained also on the boundary, that is, for $g(\theta)=R$.
Hence, for $R>R_{\text{cr}}$
\begin{align}
 & \min_{\theta\in[0,1]}f(\theta)+[g(\theta)-R]_{+}\nonumber \\
 & =\min\Bigg\{\min_{\theta\in[0,1]\colon g(\theta)\leq R}f(\theta),\min_{\theta\in[0,1]\colon g(\theta)=R}f(\theta)+g(\theta)-R\Bigg\}\\
 & =\min_{\theta\in[0,1]\colon g(\theta)\leq R}f(\theta).
\end{align}
The monotonicity properties of $f(\theta)$ and $g(\theta)$ then
imply that the solution is obtained for $\theta_{R}$ which satisfies
$g(\theta_{R})=R$.
\end{IEEEproof}
\bibliographystyle{ieeetr}
\bibliography{DNA_storage}

\end{document}